\documentclass[10pt,onecolumn,a4paper]{IEEEtran}
\usepackage{float}
\usepackage{stfloats}
\usepackage{amsmath,amsfonts,amssymb}
\usepackage{graphicx}
\usepackage{cite}
\usepackage{multirow}
\usepackage{algorithm}
\usepackage{algorithmic}
\usepackage{pifont}
\usepackage{mathtools}
\IEEEoverridecommandlockouts

\usepackage{lipsum}
\usepackage{mathtools}

\usepackage{cuted}

\newtheorem{theorem}{Theorem}
\newtheorem{remark}{Remark}

\newtheorem{lemma}{Lemma}

\title{Upper Bounds on the Capacity of 2-Layer $N$-Relay Symmetric Gaussian Network}
\author{\IEEEauthorblockN{Satyajit Thakor$^{\dagger}$ and~Syed Abbas$^{\ddagger}$}\\
\IEEEauthorblockN{School of Computing and Electrical Engineering$^{\dagger}$}\\
\IEEEauthorblockN{School of Basic Sciences$^{\ddagger}$}\\
\IEEEauthorblockN{Indian Institute of Technology Mandi}\\
\IEEEauthorblockN{email: satyajit@iitmandi.ac.in}}

\begin{document}
\maketitle
\begin{abstract}
The Gaussian parallel relay network, in which two parallel relays assist a source to convey information to a destination, was introduced by Schein and Gallager. An upper bound on the capacity can be obtained by considering broadcast cut between the source and relays and multiple access cut between relays and the destination. Niesen and Diggavi derived an upper bound for Gaussian parallel $N$-relay network by considering all other possible cuts and showed an achievability scheme that can attain rates close to the upper bound in different channel gain regimes thus establishing approximate capacity. In this paper we consider symmetric layered Gaussian relay networks in which there can be many layers of parallel relays. The channel gains for the channels between two adjacent layers are symmetrical (identical). Relays in each layer broadcast information to the relays in the next layer. For 2-layer $N$-relay Gaussian network we give upper bounds on the capacity. Our analysis reveals that for the upper bounds, joint optimization over correlation coefficients is not necessary for obtaining stronger results.
\end{abstract}

\section{Introduction}
The capacity of multihop relay networks is largely unknown. Even for a simple relay network set-up \cite{Van71} in which a relay assists a source to communicate information to a destination in addition to the direct transmission is unknown for discrete memoryless as well as Gaussian case. In \cite{SchGal00}, Schein and Gallager introduced a new network setup called Gaussian parallel relay network in which a source communicates information to the destination via two parallel relays. Cut-set outer bounds were derived and a coding scheme called amplify-and-forward (AF) was introduced. In \cite{NieDig13}, Niesen and Diggavi considered generalization of the Gaussian parallel relay network, called Gaussian diamond network, in which communication is via $N$ relays. For diamond network with symmetric channel gains an outer bound was derived.  The authors \cite{NieDig13} also proposed bursty amplify-and-forward scheme and showed constant additive and multiplicative gaps between the achievable rates for the coding scheme and the upper bound for all regimes (parameter choices). This also established an approximate capacity for the symmetric Gaussian diamond network. This results were further extended for asymmetric Gaussian diamond network by converting an asymmetric diamond network into (approximately) symmetric parallel diamond networks. 

The recent work \cite{CouOzg15} suggests that sub-linear gap (in terms of network nodes) between the cut-set bound and an achievable scheme for general relay networks is very unlikely else the cut-set bound is tight in general. However, for certain class of networks a constant factor gap may be a possibility and should be investigated. 
\begin{figure}[h]
\centering
\includegraphics[scale=.95]{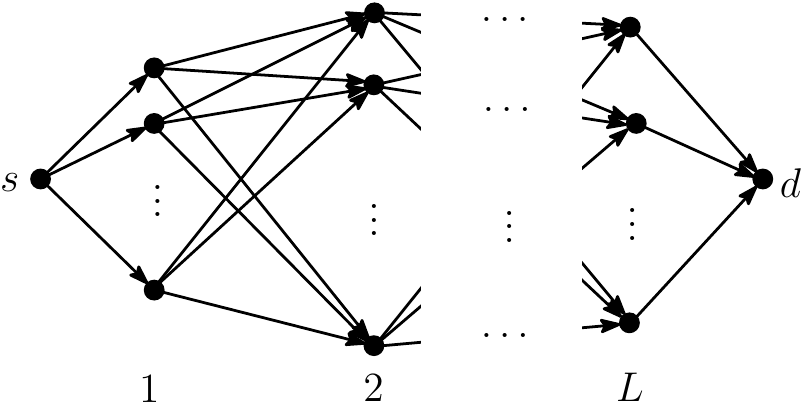}
\caption{Network with $L$ layers of parallel relays}\label{fig:Lnet0}
\end{figure}

Figure  \ref{fig:Lnet0} depicts  a generalization of the diamond network called Gaussian layered relay network. This general network has $L$ layers of parallel relays between the source and the destination. Each relay in a layer receives information from the relays of previous layers and broadcasts a function of received information to the relays in the next layer. The source node $s$ is at layer 0 and the destination node is at layer $L+1$. 
Layered deterministic relay networks were considered in \cite{AveDigTse11}. In layered networks the information flowing through different paths to a node in the networks reaches at the same time (i.e., with the same delay). This makes both upper bounding using information theoretic arguments and lower bounding using coding schemes some what easier by dropping the causality constraint for information transmission.

As noted  in \cite[p. 39]{Sch01}, some of the achievable coding schemes are seen from a better perspective once the converse results are established. In this paper, we focus on converse results (upper bounds) for the network capacity.  As a first step towards characterizing better upper bounds on the capacity of Gaussian layered relay networks (and, hopefully, establishing approximate capacity), in this paper we consider 2-layer $N$-relay Gaussian network. Diamond network is a sub-network of this network.
For this network we obtain upper bounds on the capacity which involves joint optimization over correlation parameters.
Our analysis reveals a surprising result: joint optimization over correlation parameters across different layers is not necessary to obtain a tighter cut-set type bounds \eqref{eq:boundPhi1}-\eqref{eq:boundPhi2} for 2-layer $N$-relay network. This desirable result 
may lead to characterization of an approximate capacity of layered relay networks with small multiplicative and additive gaps between the outer bounds and achievable rates for coding schemes.

The remaining part of the paper is organized as follows. Section II describes the network model and related work. Section III presents our main results. Conclusion and future directions are discussed in Section IV.

\section{Network Model}\label{sec:networkmodel}
\begin{figure}[h]
\centering
\includegraphics[scale=.51]{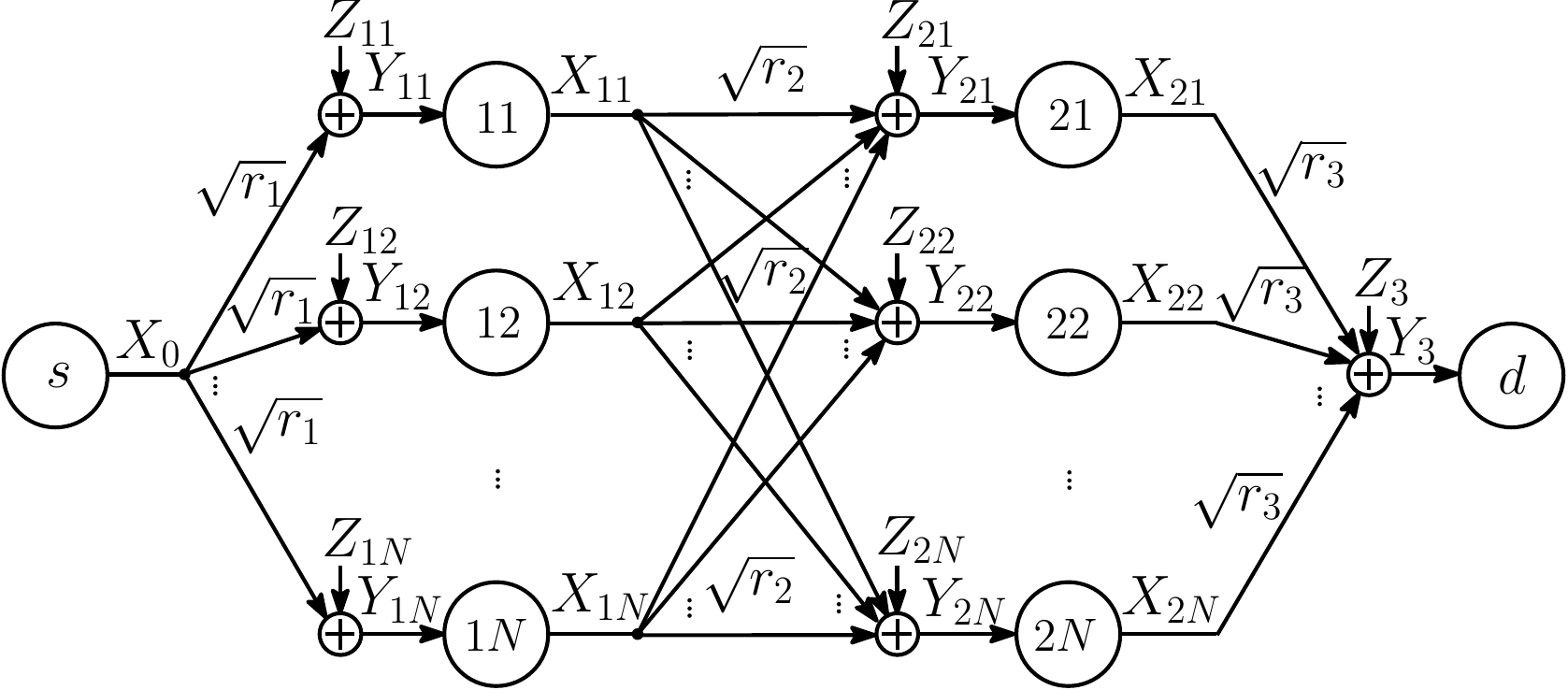}
\caption{Symmetric 2-layer $N$-realy Gaussian network}\label{fig:relay}
\end{figure}

A 2-layer $N$-relay Gaussian network with $N$ relays in each layer is shown in Figure \ref{fig:relay}. 
We adopt most notations from \cite{NieDig13}. The sets of relays are labeled $[iN]=\{i1,i2,\ldots,iN\}, i = 1,2$. In this context, $12,2k$ etc., are relays and should not be interpreted as numbers $12$ or $2 \times k$. Parameters $A_{i}, i \in I, A_{j},j \in J$ are denoted $A_{I,J}$ for simplicity. The transmitted random variables $X_{0,[1N],[2N]}$ have unit average power constraint. The channel gains $r_{1,2,3}$ are real positive constants. The channels introduce i.i.d. additive white Gaussian noise $Z_{[1N],[2N],3}$ each with zero mean and variance 1. A random variable without time index may be viewed as a generic copy.


The source message $V$ is encoded as $(X_0[t])_{t=1}^T=g_s(V)$. The input to the relay node $1j$ at time $t$ is 
$$Y_{1j}[t]=\sqrt{r_i} X_{0}[t]+Z_{1j}[t].$$

Assume that the relays incur delay of one block length $T$ (or one time instance $t$). 
In the next block interval, each of these relays broadcasts a relay code $X_{1,j}[t+T]=g_{1,j}(Y_{1,j}[t])$ to the next layer. Each relay $2j$ in layer 2 receives
$$Y_{2j}[t]= \sqrt{r_2}\sum_{l=1}^{N}X_{1l}[t]+Z_{2j}[t].$$
The coded symbols $(X_{2,j}[t+T])_{t=1}^T=g_{2,j}((Y_{2,j}[t])_{t=1}^T)$ are transmitted by relays and 
$$Y_{3}[t]= \sqrt{r_3}\sum_{l=1}^{N}X_{2l}[t]+Z_{3}[t]$$
 is received by the destination node $d$ which estimates the transmitted source message from the received message via the decoding function
 $$\hat{V}=g_d((Y_3[t])_{t=2T+1}^{3T}).$$

Note that, due to the layered structure of the network, in any block interval a relay in the second layer receives a corrupted version of the transmitted codes for the source message for the same block.  In other words $(Y_{2j}[t])_{t=T+1}^{2T}, j = 1 \ldots N$ are associated with the transmitted code block $(X_0[t])_{t=1}^T$ from the source. Also, the destination node receives a corrupted version of the transmitted codes for the source message for the same block. In other words $(Y_3[t])_{t=2T+1}^{3T}$ is associated with the transmitted code block $(X_0[t])_{t=1}^T$ from the source. As such, we assume no delay in the relay operations and drop the causality constraint for information transmission in the network. 

For the symmetric diamond network the capacity is upper bounded as follows. We refer \eqref{eq:NDbound} as the ND bound. 

\begin{theorem}[Lemma 7, \cite{NieDig13}]
For symmetric diamond network with $N$ relays and channel gains $r_1$ at source broadcast side and $r_2$ at destination mutiple-access side, the capacity
 \begin{align}
C(N,r_{1,2})&\leq \sup_{\rho_1 \in [0,1)} \min_{n \in \{0,\ldots,N\}}\frac{1}{2}\left( \log(1+(N-n)r_1) + \log (1 +\psi(n,\rho_1)r_2)\right)\label{eq:NDbound}
\end{align}
 where 
 \begin{align*}
\psi(n,\rho)= 
n \Big(1+(n-1)\rho - \frac{n(N-n)\rho^2}{1-(N-n-1)\rho} \Big).
\end{align*}
\end{theorem}

Note that, due to symmetrical gains only $N+1$ cuts need to be considered from all possible $2^N$ cuts.
This bound is a generalization of the outer bounds for Gaussian parallel 2-relay (symmetric) network in \cite{Sch01} (see also \cite{SchGal00}). In particular, \cite[Lemma 7]{NieDig13} is a generalization of the broadcast cut-set bound (2.49) and cross cut-set bounds (2.83-84) in \cite{Sch01} 
(and the simpler bound (4) in \cite{NieDig13} is a generalization of (2.49), the multiple access cut-set bound (2.58) and pure cross cut-set bound (2.68) in \cite{Sch01}).


\section{An upper bound on the capacity of 2-layer $N$-relay Gaussian network}

In this section an upper bound on the capacity of 2-layer $N$-relay network is given. 
A cut is a set of nodes $\{s\}\cup1S\cup2S$ where $1S \subseteq [1N],2S \subseteq [2N]$. The cardinalities of $1S$ and $2S$ are denoted $n$ and $m$ respectively. Also,
$$iS^c \triangleq [iN] \setminus iS, i \in \{1,2\}.$$
The three cuts $\{s\}$, $\{s\}\cup[1N]$ and $\{s\}\cup[1N]\cup[2N]$, separating each layer from the previous layer, are referred as broadcast cut, multiple broadcast-access cut and multiple access cut respectively. The set of channels in every cut in the network is a union of disjoint subset of channels in these three cuts. Accordingly, a cut is regarded to have a broadcast part, a multiple broadcast-access part and a multiple access part. We start from the cut set bound \cite{CovTho06}:
\begin{align}
C(N,r_{1,2,3}) \leq \sup_{X_{0,[1N],[2N]}} \min_{\substack{1S \subseteq [1N],\\2S \subseteq[2N]}} I(X_{0,{1S},{2S}};Y_{{1S^c},{2S^c},3}|X_{{1S^c},{2S^c}}) \label{eq:L0}
\end{align}
where, the supremum is subject to the unit transmit power constraint at the nodes. Now, 
\begin{align}
&I(X_{0,{1S},{2S}};Y_{{1S^c},{2S^c},3}|X_{{1S^c},{2S^c}}) \nonumber \\
&= h(Y_{1S^c}| X_{1S^c,2S^c}) -h(Y_{1S^c}| X_{0,[1N],[2N]})\nonumber \\
&\text{ \ \ } + h(Y_{2S^c}|Y_{1S^c},X_{1S^c,2S^c}) -h(Y_{2S^c}|Y_{1S^c}, X_{0,[1N],[2N]}) \nonumber \\
&\text{ \ \ } +h(Y_3|Y_{1S^c,2S^c}, X_{1S^c,2S^c})-h(Y_3|Y_{1S^c,2S^c}, X_{0,[1N],[2N]}) \nonumber \\
&\leq h(Y_{1S^c})+ h(Y_{2S^c}|X_{1S^c})+h(Y_3|X_{1S^c,2S^c}) \nonumber \\
&\text{ \ \ } -h(Z_{1S^c})-h(Z_{2S^c})-h(Z_3) \nonumber \\
&= h(Y_{1S^c})+ h(Y_{2S^c}|X_{1S^c})+h(Y_3|X_{1S^c,2S^c}) \nonumber \\
&\text{ \ \ } -h(Y_{1S^c}| X_{0})-h(Y_{2S^c}|X_{[1N]})-h(Y_3|X_{[1N],[2N]}) \nonumber \\
&=I(X_0;Y_{1S^c})+I(X_{1S};Y_{2S^c}|X_{1S^c})\nonumber\\
&\text{ \ \ }+I(X_{1S,2S};Y_3|X_{1S^c,2S^c}) \label{eq:L1}
\end{align}

From \eqref{eq:L0} and \eqref{eq:L1},
\begin{align}
C(N,r_{1,2,3}) \leq  \sup_{X_0,X_{[1N]},X_{[2N]}} \min_{{1S \subseteq [1N],2S \subseteq[2N]}} &\{I(X_0;Y_{1S^c})+I(X_{1S};Y_{2S^c}|X_{1S^c}) +I(X_{1S,2S};Y_3|X_{1S^c,2S^c})\}.\label{eq:L11}
\end{align}

Note that, in \eqref{eq:L1} the three conditional mutual information terms are associated with the broadcast, the multiple broadcast-access and the multiple access parts of a cut respectively. 
The first term $I(X_0;Y_{1S^c})$ in \eqref{eq:L1}, associated with the broadcast part, can be further upper bounded as
\begin{align}
I(X_0;Y_{1S^c})\leq \frac{1}{2} \log(1+|1S^c|r_1).\label{eq:L12}
\end{align}
Now, let us focus on the third term.
\begin{align}
&I(X_{1S,2S};Y_3|X_{1S^c,2S^c}) \nonumber\\
&= h(Y_3|X_{1S^c,2S^c}) - h(Y_3|X_{[1N],[2N]})  \label{eq:L2}\\
&= h\Bigl(\sqrt{r_3} \sum_{2j \in 2S} (X_{2j} - f_{2j}(X_{1S^c,2S^c})) +Z_3\Big| X_{1S^c,2S^c}\Bigr) \nonumber \\
&\text{ \ \ }  - h(Z_3)\nonumber \\
&\leq h\Bigl(\sqrt{r_3} \sum_{2j \in 2S} (X_{2j} - f_{2j}(X_{1S^c,2S^c})) +Z_3\Bigr) - h(Z_3)\label{eq:L3}
\end{align}


The equality \eqref{eq:L2} can also be written as
\begin{align}
&I(X_{1S}, X_{2S};Y_3|X_{1S^c},X_{2S^c}) \nonumber\\
&= h(Y_3|X_{2S^c}) - I(Y_3;X_{1S^c}|X_{2S^c}) - h(Y_3|X_{[1N],[2N]})\nonumber 
\end{align}
and if we ignore the term $I(Y_3;X_{1S^c}|X_{2S^c})$ then the remaining terms can be further upper bounded by the ND bound. We use the additional information $X_{1S^c}$ available across the cut in an attempt to obtain a tighter bound.

\begin{figure}[h]
\centering
\includegraphics[scale=.40]{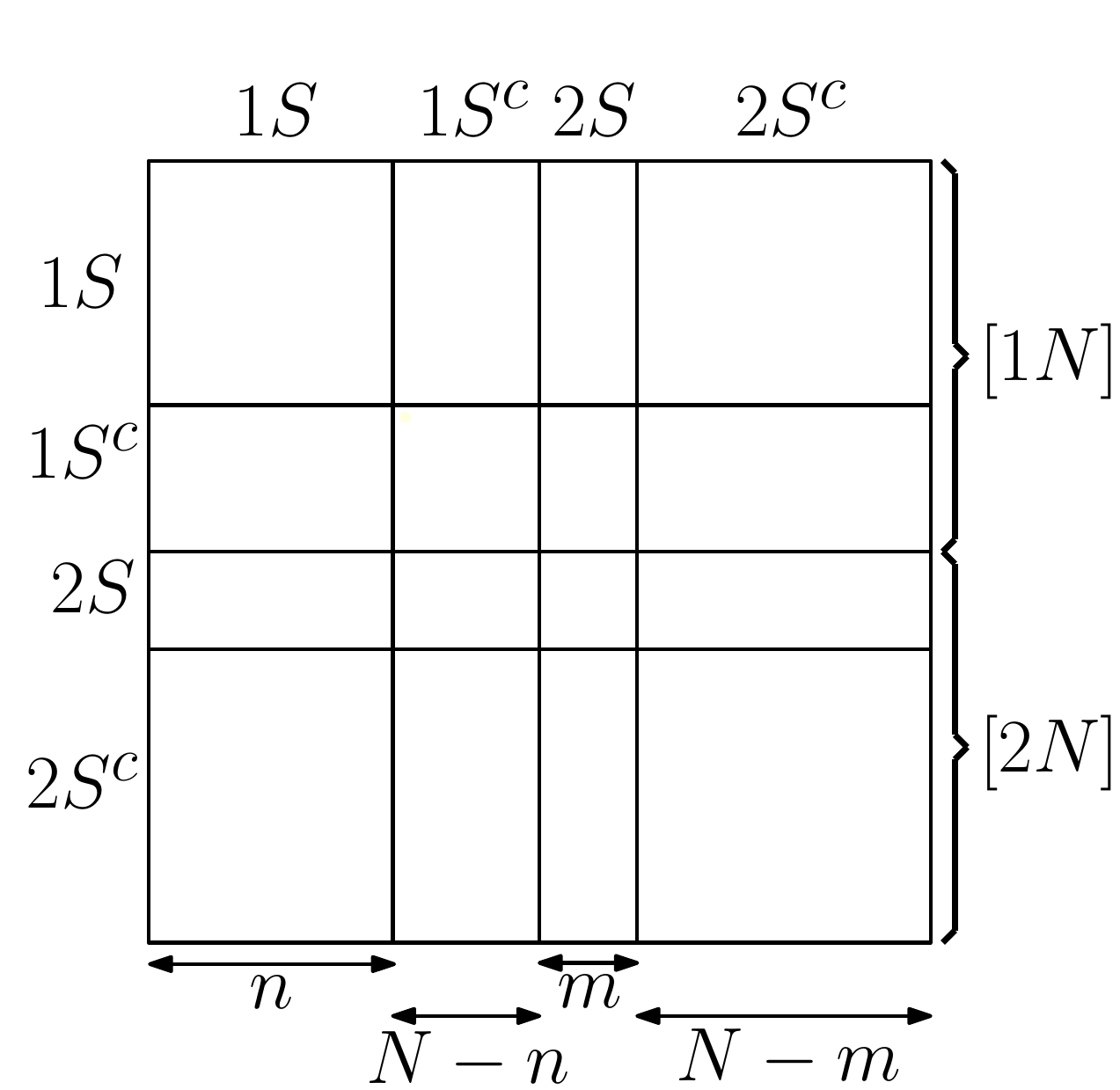}
\caption{Sketch of the matrix $\mathbf{Q}_{[1N],[2N]}$ and submatrices associated with a cut $\{s\}\cup1S\cup2S$.}\label{fig:matrix1}
\end{figure}

In \eqref{eq:L3}, 
$f_{2j}(X_{1S^c}, X_{2S^c}), 2j \in 2S$ can be any functions. But, to obtain tightest possible upper bound, it is natural to consider this function to be the ``best'' estimator of $X_{2j}$ given $X_{1S^c,2S^c}$.  If we restrict it to be the minimum mean square error (MMSE) estimator for approximating $X_{2j}$ based on $X_{1S^c},X_{2S^c}$ then the covariance matrix for the vector of random variables $(X_{2j} - f_{2j}(X_{2S^c},X_{1S^c}), 2j \in 2S)$ (note that these random variables are MMSE \cite{kai81}) can be represented as  \cite{Mui82}
\begin{align}
\mathbf{Q}_{2S\cdot 1S^c2S^c} = \mathbf{Q}_{2S,2S} - \mathbf{Q}_{2S,1S^c2S^c} \mathbf{Q}^{-}_{1S^c2S^c,1S^c2S^c} \mathbf{Q}_{1S^c2S^c,2S}
\end{align}
where, $\mathbf{Q}_{A,B}$ is the sub-matrix of the covariance matrix $\mathbf{Q}_{[1N][2N]}$ (which is, by definition, positive semidefinite) for $X_{[1N],[2N]}$ corresponding to the rows for $X_A$ and the columns for $X_B$. Also, $\mathbf{Q}^{-}_{1S^c2S^c,1S^c2S^c}$ is the Moore-Penrose generalized inverse of the matrix $\mathbf{Q}_{1S^c2S^c,1S^c2S^c}$. The matrix $\mathbf{Q}_{2S\cdot 1S^c2S^c}$ is the generalized Schur complement of $\mathbf{Q}_{1S^c2S^c,1S^c2S^c}$ in $\mathbf{Q}_{[1N][2N]}$. If $\mathbf{Q}_{1S^c2S^c,1S^c2S^c}$ is invertible, then its generalized inverse is also the inverse of the matrix and hence the generalized Schur complement reduces to the Schur complement.  The MMSE has been used to obtain upper bounds on the capacity of symmetric channel models in \cite{Tho87} as well as \cite{NieDig13}. 
A pictorial presentation of the correlation matrix $\mathbf{Q}_{[1N],[2N]}$ and submatrices associated with a generic cut $\{s\}\cup1S\cup2S$ is given in Figure \ref{fig:matrix1}.

Again, we continue with the notations in \cite{NieDig13}: $\mathbf{I}_{a}$ means the $a \times a$ identity matrix, $\mathbf{1}_{a,b}$ means $a \times b$ matrix of ones. We drop the subscripts when the dimension is clear from context. Then, the equation \eqref{eq:L3} can be further upper bounded as follows.
\begin{align}
&h\Bigl(\sqrt{r_3} \sum_{2j \in 2S} (X_{2j} - f_{2j}(X_{1S^c,2S^c})) +Z_3\Bigr) - h(Z_3)\nonumber \\
&\leq \frac{1}{2} \log (1 + r_3 \mathbf{1}^T \mathbf{Q}_{2S\cdot 1S^c2S^c}\mathbf{1})\label{eq:L4}
\end{align}

The second term $I(X_{1S};Y_{2S^c}|X_{1S^c})$ in \eqref{eq:L1}  is associated with the multiple broadcast-access part of a cut.
\begin{align}
&I(X_{1S}; Y_{2S^c}|X_{1S^c}) \nonumber\\
&= h(Y_{2S^c}|X_{1S^c}) - h(Y_{2S^c}|X_{[1N]})  \nonumber\\
&= h\Bigl(\Bigl(\sqrt{r_2} \sum_{1i \in 1S} (X_{1i} - f_{1i}(X_{1S^c})) +Z_{2j}\Bigr) : 2j \in 2S^c\Big| X_{1S^c}\Bigr)\nonumber\\
&\text{ \ \ } - h(Z_{2j}: 2j \in 2S^c) \nonumber\\
 &\leq h\Big(\Big(\sqrt{r_2} \sum_{1i \in 1S} (X_{1i} - f_{1i}(X_{1S^c})) +Z_{2j}\Big) : 2j \in 2S^c\Big) \nonumber\\
 &\text{ \ \ } - h(Z_{2j}: 2j \in 2S^c) \nonumber\\
 &\leq \frac{1}{2} |2S^c| \log (1+r_1 \mathbf{1}^T \mathbf{Q}_{1S\cdot1S^c}\mathbf{1})\label{eq:L5}
\end{align}
where $\mathbf{Q}_{1S\cdot1S^c} = \mathbf{Q}_{1S,1S}- \mathbf{Q}_{1S,1S^c}\mathbf{Q}^{-}_{1S^c,1S^c}\mathbf{Q}_{1S^c,1S}$ is the Schur complement,  $\mathbf{Q}_{1S,1S}, \mathbf{Q}_{1S,1S^c}$, $\mathbf{Q}_{1S^c,1S}$ are sub-matrices of $\mathbf{Q}_{[1N]}$ and $\mathbf{Q}^{-}_{1S^c,1S^c}$ is the Moore-Penrose generalized inverse of $\mathbf{Q}_{1S^c,1S^c}$.

Note that, ``most'' of correlation between $X_{1i}$ and $X_{1S^c}$ for all $1i \in 1S$ is ``taken care of'' by subtracting the MMSE estimator. Hence, using independence inequality for entropies (independence bound) is not the worst way of bounding the term. But, \eqref{eq:L5} is still crude since correlation among $(\sqrt{r_2} \sum_{1i \in 1S} (X_{1i} - f_{1i}(X_{1S^c})) +Z_{2j})$ for all  $2j \in 2S^c$ is not utilized in upper bounding the term.  We now obtain a different bound as follows. Let $\sqrt{r_2} \sum_{1i \in 1S} (X_{1i} - f_{1i}(X_{1S^c}))=W$ and without loss of generality, assume that $2S^c=\{21,22,\ldots, |2S^c|\}$, then
\begin{align}
&I(X_{1S}; Y_{2S^c}|X_{1S^c}) \nonumber\\
&= h(Y_{2S^c}|X_{1S^c}) - h(Y_{2S^c}|X_{[1N]})  \nonumber\\
 &\leq h((W +Z_{2j}) : 2j \in 2S^c) - h(Z_{2S^c}) \nonumber\\
 &= \sum_{2j=21}^{|2S^c|}h(W +Z_{2j}|W+Z_{2(j-1)} , \ldots, W+Z_{2|S^c|}) \nonumber\\
 &\text{ \ \ }- h(Z_{2S^c}) \nonumber\\ 
 &\leq \sum_{2j=22}^{|2S^c|}h(Z_{2j}-Z_{2(j-1)})+h(W+ Z_{21}) 
- h(Z_{2S^c}) \nonumber\\ 
 &= (|2S^c|-1)h(Z_{2j}-Z_{2j'})+h(W+ Z_{2k}) -  h(Z_{2S^c}) \nonumber\\
 &= (|2S^c|-1)\frac{1}{2}\log 4\pi e \sigma^2 +h(W+ Z_{2k}) \nonumber\\
 &\text{ \ \ }- |2S^c| \frac{1}{2}\log 2\pi e \sigma^2 \nonumber\\
 &= (|2S^c|-1)\frac{1}{2}\log 4\pi e \sigma^2 +h(W+ Z_{2k}) \nonumber\\
 &\text{ \ \ } -(|2S^c|-1) \frac{1}{2}\log 2\pi e \sigma^2-\frac{1}{2}\log 2\pi e \sigma^2 \nonumber\\
 &= \frac{|2S^c|-1}{2}+h(W+ Z_{2k})-\frac{1}{2}\log 2\pi e \sigma^2  \nonumber\\ 
 &\leq \frac{1}{2}\left(|2S^c|-1+\log (1+r_2 \mathbf{1}^T \mathbf{Q}_{1S\cdot1S^c}\mathbf{1})\right)\label{eq:L51}
\end{align}
where, $2j \neq 2j'$ and $\frac{1}{2}\log 2\pi e \sigma^2$ is the differential entropy of Gaussian distribution with variance $\sigma^2$. Note that the inequality \eqref{eq:L51} is valid only for $|2S^c|\geq 1$ (for $2S^c=\emptyset$, $I(X_{1S}; Y_{2S^c}|X_{1S^c})=0$). 
\begin{lemma}\label{lem: Lem1}
From \eqref{eq:L11}, \eqref{eq:L12}, \eqref{eq:L5}, \eqref{eq:L51}, we have \eqref{eq:boundMatrix1} for any $2S^c$ and \eqref{eq:boundMatrix2} for  nonempty $2S^c$.
\begin{align}
&C(N,r_{1,2,3}) \nonumber\\
&\leq  \sup_{\mathbf{Q}_{[1N],{[2N]}}} \min_{\substack{1S \subseteq [1N],\\2S \subseteq[2N]}}\frac{1}{2} \Big(\log(1+|1S^c|r_1)+ |2S^c| \log (1+r_2 \mathbf{1}^T \mathbf{Q}_{1S\cdot1S^c}\mathbf{1})+ \log (1 + r_3 \mathbf{1}^T \mathbf{Q}_{2S\cdot 1S^c2S^c}\mathbf{1})\Big)\label{eq:boundMatrix1}\\
&C(N,r_{1,2,3}) \nonumber\\
&\leq \sup_{\mathbf{Q}_{[1N],{[2N]}}} \min_{\substack{1S \subseteq [1N],\\2S \subseteq[2N]}} \frac{1}{2} \Big(\log(1+|1S^c|r_1)+|2S^c|-1+\log (1+r_2 \mathbf{1}^T \mathbf{Q}_{1S\cdot1S^c}\mathbf{1})+ \log (1 + r_3 \mathbf{1}^T \mathbf{Q}_{2S\cdot 1S^c2S^c}\mathbf{1})\Big)\label{eq:boundMatrix2}
\end{align}
\end{lemma}

Now we focus on the form of the covariance matrix. It  can be shown using time-sharing argument and symmetry in the network that we can restrict our attention to $\mathbf{Q}_{[1N],{[2N]}}$ of the form 
\begin{align}
\begin{pmatrix}\label{eq:L8}
  \rho_1 \mathbf{1}_{N,N}+(1-\rho_1)\mathbf{I}_{N}			& \rho_{1,2} \mathbf{1}_{N,N}\\
  \rho_{1,2} \mathbf{1}_{N,N} 	& \rho_2 \mathbf{1}_{N,N}+(1-\rho_2)\mathbf{I}_{N}
 \end{pmatrix}
\end{align} 
without loss in optimality which is equivalent to \eqref{eq:matrix1} where, the parameters $\rho_i$  represent correlation between $X_{il}$ and $X_{ik}$, $il,ik \in [iN], i\in \{1,2\}$, and $\rho_{12}$,  represents correlation between $X_{1l}$ and $X_{2k}$, $1l \in [1N],2k \in [2N]$.
\begin{align}
\begin{pmatrix}\label{eq:matrix1}
  1 			& \rho_1 	& \cdots & \rho_1 	&\rho_{12} 	& \rho_{12} 	& \cdots & \rho_{12}\\
  \rho_1 	& 1 			& \cdots & \rho_1 	&\rho_{12}	& \rho_{12} 	& \cdots & \rho_{12}\\
  \vdots  	& \vdots  & \ddots & \vdots  	&\vdots  		&	\vdots		& \ddots & \vdots\\
  \rho_1 	& \rho_1 	& \cdots & 1 			&\rho_{12}	& \rho_{12} 	& \cdots & \rho_{12}\\
  \rho_{12} 	& \rho_{12} 	& \cdots & \rho_{12}	&1 			& \rho_2 	& \cdots & \rho_2 	\\
  \rho_{12}	& \rho_{12} 	& \cdots & \rho_{12}	&\rho_2 	& 1 			& \cdots & \rho_2 	\\
  \vdots  		&	\vdots		& \ddots & \vdots		&\vdots  	& \vdots  & \ddots & \vdots  	\\
  \rho_{12}	& \rho_{12} 	& \cdots & \rho_{12}	&\rho_2 	& \rho_2 	& \cdots & 1 			
 \end{pmatrix}
\end{align} 
\subsubsection*{Sketch of proof}First fix a cut under consideration and now suppose a matrix of some another form is optimal. Note that all matrices obtained by permutation mappings $\sigma_1: [1N]\longrightarrow[1N]$ and $\sigma_2: [2N]\longrightarrow[2N]$ are also optimal due to symmetry. Now, time sharing between all such $(N!)^2$ matrices is also optimal. By time-sharing all the matrices for equal time duration, the matrix for such time sharing scheme has the form  \eqref{eq:matrix1}.

Positive semidefinite property of a covariance matrix restricts the range of $\rho_1, \rho_2, \rho_{12}$ as follows (see Appendix \ref{ap:rangeRhos} for details).
\begin{align*}
\rho_1,\rho_2 &\in [-1,1]\\
\rho_{12} &\in \left[-1,\min\left\{\frac{1+(N-1)\rho_1}{N},\frac{1+(N-1)\rho_2}{N}\right\}\right] \triangleq \zeta
\end{align*}
Note that $\mathbf{1}^T \mathbf{Q}_{1S\cdot1S^c}\mathbf{1} = \psi(n,\rho_1)$ is already derived in \cite{NieDig13} and the valid range for $\rho_1$ is also given.  Now we turn our attention to deriving expression for $\mathbf{1}^T \mathbf{Q}_{2S\cdot 1S^c2S^c}\mathbf{1}$. This involves finding inverse of sub-matrices associated with cuts (see Appendix \ref{app:matrices} for details) and the Schur complement. We derive the expression for Schur complement as follows (see Appendix \ref{ap:schur} for details). We derive the expression for $\mathbf{1}^T \mathbf{Q}_{2S\cdot 1S^c2S^c}\mathbf{1}$ and is given in \eqref{eq:FnPhi} (see Appendix \ref{ap:schurMutiplication}).
\begin{figure*}[!ht]
 \begin{align}
 \mathbf{1}^T \mathbf{Q}_{2S\cdot 1S^c2S^c}\mathbf{1}=
\phi(n,m,\rho_{1,2,{12}})&=m\big(1+(m-1)\rho_2 - m(N-n)\rho_{12}^2 (x+(N-n-1)y)\nonumber\\
&\text{ \ \ \ \  \ \ }- m(N-n)^2(N-m) \rho_{12}^4 (e+ (N-m-1)f)(x+(N-n-1)y)^2\nonumber\\
&\text{ \ \ \ \  \ \ }+2m(N-m)(N-n)\rho_2 \rho_{12}^2(e+(N-m-1)f)(x+(N-n-1)y) \nonumber\\
&\text{ \ \ \ \  \ \ }-m(N-m)\rho_{2}^2(e + (N-m-1)f) \big)\label{eq:FnPhi}\\
\text{where }x+(N-n-1)y&= \frac{1}{1+(N-n-1)\rho_1}\nonumber\\
\text{and } e+(N-m-1)f
&=\frac{(1+(N-n-1)\rho_1)}{(1+(N-n-1)\rho_1)(1+(N-m-1)\rho_2)-(N-m)(N-n)\rho_{12}^2}.\nonumber
\end{align}
\hrulefill
\end{figure*}

Note that, 
If $1S=2S=\emptyset$, then (we take the convention $\mathbf{Q}_{\emptyset,A}=\mathbf{0}_{1,|A|}$, $\mathbf{Q}_{A,\emptyset}=\mathbf{0}_{|A|,1}$ and  $\mathbf{Q}_{\emptyset,\emptyset}=0$)
\begin{align}
\mathbf{1}^T \mathbf{Q}_{2S\cdot 1S^c2S^c}\mathbf{1}=0. \label{eq:specialcase1}
\end{align}

Since the derivation of $\phi(\cdot)$ involves (or assumes) inverting $\mathbf{Q}_{1S^c2S^c,1S^c2S^c}$, we cannot use the function $\phi(\cdot)$ as a representation of $\mathbf{1}^T \mathbf{Q}_{2S\cdot 1S^c2S^c}\mathbf{1}$ when for certain values of correlation coefficients the submatrices $\mathbf{Q}_{1S^c,1S^c}$ and $\mathbf{Q}_{1S^c2S^c,1S^c2S^c}$  corresponding to cuts are not invertible. In particular, for the following cases the matrix $\mathbf{Q}_{1S^c2S^c,1S^c2S^c}$ is not invertible and we use Moore-Penrose generalized inverse to find generalized Schur complement.  

\begin{itemize}
\item At the other extreme (compared to the situation in \eqref{eq:specialcase1}) if $1S\cup2S=[1N]\cup[2N]$ then
\begin{align}
\mathbf{1}^T \mathbf{Q}_{2S\cdot 1S^c2S^c}\mathbf{1}=N(1+(N-1)\rho_2).\label{eq:specialcase2}
\end{align}
\item If $1S \subsetneq [1N], 2S = [2N]$ and $\rho_1=1$ then
\begin{align}
\mathbf{1}^T \mathbf{Q}_{2S\cdot 1S^c2S^c}\mathbf{1}=N(1+(N-1)\rho_2-N\rho_{12}^2). \label{eq:specialcase3}
\end{align}
\item If $2S \subsetneq [2N], 1S = [1N]$ and $\rho_2=1$ then 
\begin{align}
\mathbf{1}^T \mathbf{Q}_{2S\cdot 1S^c2S^c}\mathbf{1}=0
\end{align}
\item If $\emptyset \neq 1S\cup2S \subsetneq [1N]\cup[2N]$ and $\rho_1=\rho_2=\rho_{12}=1$, then
\begin{align}
\mathbf{1}^T \mathbf{Q}_{2S\cdot 1S^c2S^c}\mathbf{1}=0.\label{eq:specialcase5}
\end{align}
\end{itemize}



Now, for these values the function $\phi(\cdot)$ is undefined:
\begin{align}
\rho_{1,2,12}&=1 \label{eq:undef1}\\
\rho_1&=\frac{-1}{N-1},n=0 \label{eq:undef2}
\end{align} 
Note that $\phi(\cdot)$ seems undefiend if we put the values $n=N$ and $\rho_1=1$ simulteneously. But, if we evaluate $\phi(\cdot)$ case by case (cut by cut) basis, i.e., first fixing a cut and thus a value of $n$ (e.g., $N$) and evaluating the function for this value, and then evaluate the function with the value of $\rho_1$ (e.g., 1) then, $\phi(\cdot)$ is in fact defined. Similar is the case with the values (1) $\rho_2=1$ and $m=N$ and, (2) $\rho_2=\frac{-1}{N-1}$ and $m=0$.

But, for the parameter values \eqref{eq:undef1}, we have
\begin{align}
\lim_{\rho_{12} \uparrow 1}\phi(n,m,\rho_{1}=1,\rho_{2}=1,\rho_{12})&=\mathbf{1}^T \mathbf{Q}_{2S\cdot 1S^c2S^c}\mathbf{1}_{\mid \rho_{1,2,12}=1}\label{eq:limit1}
\end{align} 
which follows from \eqref{eq:specialcase1}-\eqref{eq:specialcase5}. 
Now, note that for the situation \eqref{eq:undef2}, $\phi(\cdot)$ is undefined (the limit is positive infinity) only when $n=0$ and $m>0$. But $n=0,m=0$ implies broadcast cut and $n=0, m>0$ is practically meaningless since the channels in the broadcast cut already separates $s$ from $d$ (and hence there is no need to add more channels to the cut). As such, we can assume that if $n=0$ then $m=0$ and thus avoid the situations when $\phi(\cdot)$ is undefined. Actually, it may be assumed that $n\geq m$ without loss of generality, but this assumption is unnecessary (does not lead to undefined function) for most cases except when $m=0$.


There may be still more parameter values for which the matrix is not invertible. For examining this values we first make rank preserving transformation of the matrix \eqref{eq:matrix1}, as shown in \eqref{eq:matrix2}. Then we check the paprameter values for which the submatrices $\mathbf{Q}_{1S^c,1S^c}$ and $\mathbf{Q}_{1S^c2S^c,1S^c2S^c}$ are non-invertible.
\begin{align}
\begin{pmatrix}\label{eq:matrix2}
  1 			& \rho_1 	& \rho_1	&\cdots & \rho_1 &\rho_{12} 	& \rho_{12} 	& \rho_{12} 	& \cdots & \rho_{12}\\
  \rho_1-1 	& 1-\rho_1 & 0			&\cdots & 0 		  &0				& 0 				& 0 				& \cdots & 0\\
  \rho_1-1 	& 0 			& 1-\rho_1	&\cdots & 0 		  &0				& 0 				& 0 				& \cdots & 0\\
  \vdots  	& \vdots		& \vdots  	& \ddots& \vdots  &\vdots		&\vdots  		&	\vdots		& \ddots & \vdots\\
  \rho_1-1 	& 0			&0 			& \cdots& 1-\rho_1 &0				&0				& 0 				& \cdots & 0\\
  \rho_1-1 	& 0			&0 			& \cdots& 1-\rho_1 &0				&0				& 0 				& \cdots & 0\\
   0				& 0 				& 0 				& \cdots & 0&1-\rho_2 	& \cdots 			& 0	&0 & \rho_2-1 		  \\
 \vdots  	& \vdots		& \vdots  	& \ddots& \vdots  &\vdots		&\vdots  		&	\vdots		& \ddots & \vdots\\
  0				& 0 				& 0 				& \cdots & 0&0 	& \cdots 			& 1-\rho_2	&0 & \rho_2-1 		  \\
  0				& 0 			& 0 			& \cdots & 0		  & 0	& \cdots& 0 			&1-\rho_2 & \rho_2-1 		  \\
   \rho_{12} & \rho_{12}& \rho_{12}&\cdots&\rho_{12}	&\rho_2 &\cdots 	& \rho_2 		& \rho_2		& 1
 \end{pmatrix}
\end{align} 

Now, note that the submatrices  $\mathbf{Q}_{1S^c2S^c,1S^c2S^c}$ associated with this new matrix \eqref{eq:matrix2} are not invertible (that is, the submatrices are not full-rank) when 
$\rho_{1}=1,\rho_2=1, \rho_{12}=-1$ or when $\rho_{12}=-1, n=N-1, m=N-1$. 
For the first case the limit of the function is as follows.
\begin{align}
\lim_{\rho_{12} \downarrow -1}\phi(n,m,\rho_{1}=1,\rho_2=1,\rho_{12})&=0
\end{align}
For the second case 
$$\mathbf{Q}_{1S^c2S^c,1S^c2S^c}=\begin{pmatrix}
 1 & -1 \\
 -1 & 1
 \end{pmatrix}$$
we have
\begin{align}
\mathbf{1}^T \mathbf{Q}_{2S\cdot 1S^c2S^c}\mathbf{1}&= (N-1)\Big(1+(N-2)\rho_2-\frac{(N-1)(1+\rho_{2})^2}{4}\Big)\label{eq:specialcase}\\
&\triangleq \mu (\rho_2)
\end{align}
whereas $\phi(\cdot) \rightarrow -\infty$ as $\rho_{12} \rightarrow -1$ except for $N=1$ or $\rho_2=1$ for which the limit and \eqref{eq:specialcase} are zero. Hence, for this special case, we must rely on  \eqref{eq:specialcase} to evaluate the bound. Also, the right hand side of \eqref{eq:specialcase} is positive when these two conditions are satisfied: $\rho_2 < 1$ and $\rho_2> \frac{N-5}{N-1}$.

Combining these results with \eqref{eq:boundMatrix1}-\eqref{eq:boundMatrix2} render the following bounds.
\begin{lemma}\label{lem:boundsPhi}
The capacity of 2-layer $N$-relay Gaussian network is upper bounded as 
\begin{align}
C(N,r_{1,2,3}) &\leq  \sup_{\substack{\rho_1,\rho_2 \in [-1,1),\\\rho_{12} \in \zeta}} \min_{n,m \in \{0,\ldots,N\}}\frac{1}{2} \Big(\log(1+(N-n)r_1)\nonumber \\
&\text{ \ \ }+(N-m) \log (1+\psi(n,\rho_{1})r_2)\nonumber\\
&\text{ \ \ }+ \log (1 + \phi(n,m,\rho_{1,2,{12}})r_3)\Big)\label{eq:boundPhi1}
\end{align}
for $N-m \geq 0$ and
\begin{align}
C(N,r_{1,2,3}) &\leq  \sup_{\substack{\rho_1,\rho_2 \in [-1,1),\\\rho_{12} \in \zeta}} \min_{n,m \in \{0,\ldots,N\}} \frac{1}{2} \Big(\log(1+(N-n)r_1)\nonumber\\
&\text{ \ \ }+N-m-1+\log (1+\psi(n,\rho_{1})r_2)\nonumber\\
&\text{ \ \ }+\log (1 + \phi(n,m,\rho_{1,2,{12}})r_3)\Big).\label{eq:boundPhi2}
\end{align}
for $N-m \geq 1$ where, if $n=0$ then $m=0$ for both \eqref{eq:boundPhi1} and \eqref{eq:boundPhi2} and if $n=N-1, m=N-1$ and $\rho_{12}=-1$ then $\rho_2=1$. For the special case when $n=N-1, m=N-1,\rho_{12}=-1$ and $\rho_{2}<1$ we have  
\begin{align}
C(N,r_{1,2,3}) &\leq  \frac{1}{2} \Big(\log(1+r_1)\nonumber \\
&\text{ \ \ }+ \sup_{\rho_1 \in [-1,1)}\log(1+\psi(n,\rho_{1})r_2)\nonumber\\
&\text{ \ \ }+ \sup_{\rho_2 \in [\max \{-1,\frac{N-5}{N-1}\},1)}\log (1 + \mu(\rho_2) r_3)\Big).
\end{align}
\end{lemma}
\begin{remark}
By letting $\rho_1,\rho_2 \in [-1,1)$, the possibility of $1 \in \zeta$ is already eliminated. In other words, $\rho_1 \neq 1$ or $\rho_2 \neq 1$ implies $\rho_{12}\neq 1$. 
\end{remark}

Let, $(n^*,m^*)$ be a pair from $\{0,\ldots,N\}\times\{0,\ldots,N\}$ such that it minimizes the quantity in \eqref{eq:boundPhi1} or in \eqref{eq:boundPhi2}. We call such a pair as \textit{minimizer pair}. 

Note that if we put $\rho_{12}=0$ in $\phi(n^*,m^*,\rho_{1,2,{12}})$ then it reduces to $\psi(m^*,\rho_{2})$. Specifically,
\begin{align*}
\sup_{\rho_{1,2} \in [-1,1), \rho_{12} =0}\phi(n^*, m^*,\rho_{1,2,{12}})&\stackrel{(a)}{=}\sup_{\rho_2 \in [-\frac{1}{N-1}, 1)}\psi(m^*,\rho_2)\\
& \stackrel{(b)}{=} \sup_{\rho_2 \in [0, 1)}\psi(m^*,\rho_2)
\end{align*}
where the lower bound for $\rho_2$ in (a) follows from the range $\zeta$ for $\rho_{12}$ and (b) follows from \cite[Lemma 7]{NieDig13}.
 Moreover, $\rho_{12}=0$ is one of the valid values to optimize the function $\phi(n,m,\rho_{1,2,{12}})$. The parameter value $\rho_{12}=0$ also enforces the lower bound $\frac{-1}{N-1}$ for $\rho_1$. 
 Thus, 
\begin{lemma}\label{lem:jointOpt}
For 2-layer $N$-relay networks, joint maximization over the parameters $\rho_{1,2,{12}}$ in \eqref{eq:boundPhi1}-\eqref{eq:boundPhi2} does not render a better bound.
\end{lemma} 
\begin{remark}
These observations may be of interest - The function $\phi(n,m,\rho_{1,2,{12}})$ is even in $\rho_{12}$ suggesting that its values may be restricted to non-negative (or non-positive) range (without loss in optimization). Moreover, $\rho_{12}=0$ is a critical point and is a maxima (see Appendix \ref{ap:criticalNmaxima}).
\end{remark}
Now we state upper bounds on the capacity which follows from Lemma \ref{lem:boundsPhi} and \ref{lem:jointOpt}.
\begin{theorem}
For 2-layer $N$-relay networks, the capacity is upper bounded as
\begin{align}
C(N,r_{1,2,3}) &\leq \frac{1}{2}\Big( \log(1+(N-n^*)r_1) \nonumber\\
&{ \ \ }+ (N-m^*) \sup_{\rho_1\in [0,1)} \log(1+\psi(n^*,\rho_1) r_2) \nonumber\\
&{ \ \ }+ \sup_{\rho_2 \in [0,1)} \log(1+\psi(m^*,\rho_2) r_3)\Big)\label{eq:bound2lN-1}
\end{align}
for $N-m^* \geq 0$ and
\begin{align}
C(N,r_{1,2,3}) 
&\leq \frac{1}{2}\Big( \log(1+(N-n^*)r_1)+ N-m^*-1 \nonumber\\
&{ \ \ } + \sup_{\rho_1 \in [0,1)} \log(1+\psi(n^*,\rho_1) r_2)\nonumber\\
&{ \ \ }+ \sup_{\rho_2 \in [0,1)} \log(1+\psi(m^*,\rho_2) r_3)\Big)\label{eq:bound2lN-2}
\end{align}
for $N-m\geq 1$ and for all minimizer pairs $(n^*,m^*)$.
\end{theorem}


Both the bounds, \eqref{eq:bound2lN-1} and \eqref{eq:bound2lN-2}, are equal when $\sup_{\rho_1\in [0,1)} \log(1+\psi(n*,\rho_1) r_2)=1$ or when $N-m^*=1$. For $N-m^* \geq 2$, the bound \eqref{eq:bound2lN-1} is tighter in the range $\sup_{\rho_1\in [0,1)} \log(1+\psi(n^*,\rho_1) r_2)<1$ and the bound \eqref{eq:bound2lN-2} is tighter in the range $\sup_{\rho_1\in [0,1)} \log(1+\psi(n^*,\rho_1) r_2)>1$.


\section{Conclusion and future directions}
In this work we derived upper bounds on 2-layer $N$-relay Gaussian network and showed that joint optimization of correlation coefficients is not necessary for evaluating a better bound. We are investigating generalization of the upper bounds for general $L$-layer symmetric networks. Other directions of future research are to (1) analyze and compare achievable rates for existing and new schemes, and to characterize the gap between the upper and lower bounds on the capacity, (2) investigate whether for $L$-layer $N$-relay networks the joint optimization of correlation coefficient is necessary, and (3) extend the bounds for the case of asymmetric gains.

In general, the joint optimization across different layers (in non-layered) networks may render tighter bound. It would be interesting to characterize a class of networks for which the joint optimization leads to a better cut-set bound.

\bibliographystyle{ieeetr}
\bibliography{network}

\begin{appendices}

\section{Range of correlation coefficients}\label{ap:rangeRhos}
Now we find the range of $\rho_1, \rho_2, \rho_{12}$ for the matrix to be positive semidefinite.
For matrices $\mathbf{A},\mathbf{B},\mathbf{C}$ of the same dimension,
\begin{align}
\textrm{det} \begin{pmatrix}
  \mathbf{A}  	& \mathbf{C}\\
  \mathbf{C} 	& \mathbf{B}
 \end{pmatrix}
 = \textrm{det}(\mathbf{AB}-\mathbf{C}^2)
\end{align} 
Letting $\mathbf{A}=\rho_1 \mathbf{1}_{N,N}+(1-\rho_1)\mathbf{I}_{N}$, $\mathbf{B}=\rho_2 \mathbf{1}_{N,N}+(1-\rho_2)\mathbf{I}_{N}$ and $\mathbf{C}=\rho_{12} \mathbf{1}_{N,N}$, we get
\begin{align*}
&\textrm{det}(\mathbf{Q}_{[1N],{[2N]}}-\lambda\mathbf{I})\\
&= \left[(1-\lambda)^2+(N-1)(1-\lambda)\rho_1 +(N-1)(1-\lambda)\rho_2+(N-1)^2\rho_1\rho_2-N^2\rho_{12}^2\right]\\
&{\hspace{3cm}}((1-\lambda)^2-(1-\lambda)\rho_1-(1-\lambda)\rho_2+\rho_1\rho_2)^{N-1}
\end{align*}
The eigenvalues are
\begin{align*}
\lambda&=1-\rho_{1}\\
&=1-\rho_{2}\\
&=1-N\rho_{12}+{(N-1)\rho_1}\\
&=1-N\rho_{12}+{(N-1)\rho_2}
\end{align*}
Hence, $\rho_1,\rho_2$ must be less than 1 and $\rho_{12} \leq \min\{\frac{1+(N-1)\rho_1}{N},\frac{1+(N-1)\rho_2}{N}\}$  for the matrix to be a positive semidefinite. Thus the range of the correlation coefficients is $\rho_1,\rho_2 \in [-1,1]$ and $\rho_{12} \in [-1,\min\{\frac{1+(N-1)\rho_1}{N},\frac{1+(N-1)\rho_2}{N}\}]$.

\section{Finding inverse of sub-matrices associated with cuts to find Schur complements}\label{app:matrices}

The following is the inverse of a sub-matrix of $\mathbf{Q}_{[1N],{[2N]}}$ corresponding to a cut where $\mathbf{A}=\rho_1 \mathbf{1}_{n,n}+(1-\rho_1)\mathbf{I}_{n}$ is $n\times n$ and $n\in \{0,1,\ldots,N\}$,   $\mathbf{B}=\rho_{12} \mathbf{1}_{n,m}$ is $n\times m$ with entries $\rho_{12}$, $\mathbf{C}=\rho_{12} \mathbf{1}_{m,n}$ is $m\times n$ with entries $\rho_{12}$ and $\mathbf{D}=\rho_2 \mathbf{1}_{m,m}+(1-\rho_2)\mathbf{I}_{m}$ is $m\times m$ where $m \in \{0,1,\ldots,N\}$.
\begin{align}
\begin{pmatrix} \mathbf{A} & \mathbf{B} \\ \mathbf{C} & \mathbf{D} \end{pmatrix}^{-1} = \begin{pmatrix} \mathbf{A}^{-1}+\mathbf{A}^{-1}\mathbf{B}(\mathbf{D}-\mathbf{CA}^{-1}\mathbf{B})^{-1}\mathbf{CA}^{-1} & -\mathbf{A}^{-1}\mathbf{B}(\mathbf{D}-\mathbf{CA}^{-1}\mathbf{B})^{-1} \\ -(\mathbf{D}-\mathbf{CA}^{-1}\mathbf{B})^{-1}\mathbf{CA}^{-1} & (\mathbf{D}-\mathbf{CA}^{-1}\mathbf{B})^{-1} \end{pmatrix}
\end{align}
\begin{remark}
In this section calculations are with respect to the sets sizes $|1S|=N-n$ and $|2S|=N-m$. Note that letting $|1S|=n$ and $|2S|=m$ does not change the set of bounds. 
\end{remark}
\begin{align}
\mathbf{A}^{-1}_{n \times n}=\begin{pmatrix}
  1 			& \rho_1 	& \cdots & \rho_1\\
  \rho_1 	& 1 			& \cdots & \rho_1\\
  \vdots  	& \vdots  & \ddots & \vdots \\
  \rho_1 	& \rho_1 	& \cdots & 1
 \end{pmatrix}^{-1}=\begin{pmatrix}
  \frac{(n-2)\rho_1+1}{-(n-1)\rho_1^2+(n-2)\rho_1+1} 			& \frac{-\rho_1}{-(n-1)\rho_1^2+(n-2)\rho_1+1} 				& \cdots \\
  \frac{-\rho_1}{-(n-1)\rho_1^2+(n-2)\rho_1+1} 						& \frac{(n-2)\rho_1+1}{-(n-1)\rho_1^2+(n-2)\rho_1+1} 	& \cdots \\
  \vdots  	& \vdots  & \ddots 	
 \end{pmatrix}=\begin{pmatrix}
  x& y& \cdots &y\\
  y& x& \cdots &y\\
  \vdots  	& \vdots  & \ddots 	& \vdots\\
    y& y& \cdots &x\\
 \end{pmatrix}
\end{align} 

\begin{align}
\mathbf{C}\mathbf{A}^{-1}\mathbf{B}_{m \times m}=\begin{pmatrix}
  n\rho_{12} ((n-1)\rho_{12}y+\rho_{12}x)		& n\rho_{12} ((n-1)\rho_{12}y+\rho_{12}x)		& \cdots \\
  n\rho_{12} ((n-1)\rho_{12}y+\rho_{12}x)		& n\rho_{12} ((n-1)\rho_{12}y+\rho_{12}x)	 	& \cdots \\
  \vdots  	& \vdots  & \ddots 	
 \end{pmatrix}=\begin{pmatrix}
  z& z& \cdots &z\\
  z& z& \cdots &z\\
  \vdots  	& \vdots  & \ddots 	& \vdots\\
    z& z& \cdots &z\\
 \end{pmatrix}
\end{align} 
 where $\mathbf{C}\mathbf{A}^{-1}\mathbf{B}$ is a $m \times m$ matrix.
\begin{align}
\mathbf{D}-\mathbf{C}\mathbf{A}^{-1}\mathbf{B}_{m \times m}=\begin{pmatrix}
  1-z& \rho_2-z& \cdots &\rho_2-z\\
  \rho_2-z& 1-z& \cdots &\rho_2-z\\
  \vdots  	& \vdots  & \ddots 	& \vdots\\
    \rho_2-z& \rho_2-z& \cdots &1-z\\
 \end{pmatrix}=\begin{pmatrix}
  u& v& \cdots &v\\
  v& u& \cdots &v\\
  \vdots  	& \vdots  & \ddots 	& \vdots\\
    v& v& \cdots &u\\
 \end{pmatrix}
\end{align}

\begin{align}
(\mathbf{D}-\mathbf{C}\mathbf{A}^{-1}\mathbf{B})^{-1}_{m \times m}=\begin{pmatrix}
  \frac{(m-2)v+u}{-(m-1)v^2+(m-2)uv+u^2} 			& \frac{-v}{-(m-1)v^2+(m-2)uv+u^2} 				& \cdots \\
  \frac{-v}{-(m-1)v^2+(m-2)uv+u^2}						&\frac{(m-2)v+u}{-(m-1)v^2+(m-2)uv+u^2}  	& \cdots \\
  \vdots  	& \vdots  & \ddots 	
 \end{pmatrix}=\begin{pmatrix}
  e& f& \cdots &f\\
  f& e& \cdots &f\\
  \vdots  	& \vdots  & \ddots 	& \vdots\\
    f& f& \cdots &e\\
 \end{pmatrix}
\end{align} 

\begin{align}
&\mathbf{B}(\mathbf{D}-\mathbf{C}\mathbf{A}^{-1}\mathbf{B})^{-1}\mathbf{C}_{n \times n}\\
&=\begin{pmatrix}
  m\rho_{12} ((m-1)\rho_{12}f+\rho_{12}e)		& m\rho_{12} ((m-1)\rho_{12}f+\rho_{12}e)		& \cdots \\
  m\rho_{12} ((m-1)\rho_{12}f+\rho_{12}e)		& m\rho_{12} ((m-1)\rho_{12}f+\rho_{12}e)	 	& \cdots \\
  \vdots  	& \vdots  & \ddots 	
 \end{pmatrix}=\begin{pmatrix}
  w& w& \cdots &w\\
  w& w& \cdots &w\\
  \vdots  	& \vdots  & \ddots 	& \vdots\\
    w& w& \cdots &w\\
 \end{pmatrix}
\end{align}

\begin{align}
&\mathbf{A}^{-1}\mathbf{B}(\mathbf{D}-\mathbf{CA}^{-1}\mathbf{B})^{-1}\mathbf{CA}^{-1}_{n \times n}\\
&=\begin{pmatrix}
  x& y& \cdots &y\\
  y& x& \cdots &y\\
  \vdots  	& \vdots  & \ddots 	& \vdots\\
    y& y& \cdots &x\\
 \end{pmatrix} \times \begin{pmatrix}
  w& w& \cdots &w\\
  w& w& \cdots &w\\
  \vdots  	& \vdots  & \ddots 	& \vdots\\
    w& w& \cdots &w\\
 \end{pmatrix}\times \begin{pmatrix}
  x& y& \cdots &y\\
  y& x& \cdots &y\\
  \vdots  	& \vdots  & \ddots 	& \vdots\\
    y& y& \cdots &x\\
 \end{pmatrix} \\
& =\begin{pmatrix}
  x& y& \cdots &y\\
  y& x& \cdots &y\\
  \vdots  	& \vdots  & \ddots 	& \vdots\\
    y& y& \cdots &x\\
 \end{pmatrix} \times \begin{pmatrix}
  w(x+(n-1)y)& w(x+(n-1)y)& \cdots &w(x+(n-1)y)\\
  w(x+(n-1)y)& w(x+(n-1)y)& \cdots &w(x+(n-1)y)\\
  \vdots  	& \vdots  & \ddots 	& \vdots\\
    w(x+(n-1)y)& w(x+(n-1)y)& \cdots &w(x+(n-1)y)\\
 \end{pmatrix}\\
& =\begin{pmatrix}
  w(x+(n-1)y)^2& w(x+(n-1)y)^2& \cdots &w(x+(n-1)y)^2\\
  w(x+(n-1)y)^2& w(x+(n-1)y)^2& \cdots &w(x+(n-1)y)^2\\
  \vdots  	& \vdots  & \ddots 	& \vdots\\
    w(x+(n-1)y)^2& w(x+(n-1)y)^2& \cdots &w(x+(n-1)y)^2\\
 \end{pmatrix}
\end{align}

\begin{align}
&\mathbf{A}^{-1}+\mathbf{A}^{-1}\mathbf{B}(\mathbf{D}-\mathbf{CA}^{-1}\mathbf{B})^{-1}\mathbf{CA}^{-1}_{n \times n}\\
&=\begin{pmatrix}
  x+w(x+(n-1)y)^2& y+w(x+(n-1)y)^2& \cdots &y+w(x+(n-1)y)^2\\
  y+w(x+(n-1)y)^2& x+w(x+(n-1)y)^2& \cdots &y+w(x+(n-1)y)^2\\
  \vdots  	& \vdots  & \ddots 	& \vdots\\
    y+w(x+(n-1)y)^2& y+w(x+(n-1)y)^2& \cdots &x+w(x+(n-1)y)^2\\
 \end{pmatrix}\\
 &=\begin{pmatrix}
  \alpha_{A}& \beta_{A}& \cdots &\beta_{A}\\
  \beta_{A}& \alpha_{A}& \cdots &\beta_{A}\\
  \vdots  	& \vdots  & \ddots 	& \vdots\\
    \beta_{A}& \beta_{A}& \cdots &\alpha_{A}\\
 \end{pmatrix}
\end{align} 

\begin{align}
&\mathbf{B}(\mathbf{D}-\mathbf{CA}^{-1}\mathbf{B})^{-1}_{n \times m}\\
&=\begin{pmatrix}
  \rho_{12}((m-1)f+e)		&\rho_{12}((m-1)f+e)		&  \cdots \\
  \vdots  	& \vdots  &\ddots	
 \end{pmatrix}=\begin{pmatrix}
  h& h& \cdots &h\\
  h& h& \cdots &h\\
  \vdots  	& \vdots  & \ddots 	& \vdots\\
    h& h& \cdots &h\\
 \end{pmatrix}
\end{align} 

\begin{align}
-\mathbf{A}^{-1}\mathbf{B}(\mathbf{D}-\mathbf{CA}^{-1}\mathbf{B})^{-1}_{n \times m}\\
&=-\begin{pmatrix}
  h((n-1)x+y)		&h((n-1)x+y)		&  \cdots \\
  \vdots  	& \vdots  &\ddots	
 \end{pmatrix}=\begin{pmatrix}
  \alpha_{B}& \alpha_{B}& \cdots &\alpha_{B}\\
  \alpha_{B}& \alpha_{B}& \cdots &\alpha_{B}\\
  \vdots  	& \vdots  & \ddots 	& \vdots\\
  \alpha_{B}& \alpha_{B}& \cdots &\alpha_{B}\\
 \end{pmatrix}
\end{align}

\begin{align}
\mathbf{CA}^{-1}_{m \times n}=\begin{pmatrix}
  \rho_{12}((n-1)y+x)		&\rho_{12}((n-1)y+x)		&  \cdots \\
  \vdots  	& \vdots  &\ddots	
 \end{pmatrix}=\begin{pmatrix}
  i& i& \cdots &i\\
  i& i& \cdots &i\\
  \vdots  	& \vdots  & \ddots 	& \vdots\\
    i& i& \cdots &i\\
 \end{pmatrix}
\end{align} 

\begin{align}
-(\mathbf{D}-\mathbf{CA}^{-1}\mathbf{B})^{-1}\mathbf{CA}^{-1}_{m \times n}
=-\begin{pmatrix}
  i((m-1)f+e)		&i((m-1)f+e)		&  \cdots \\
  \vdots  	& \vdots  &\ddots	
 \end{pmatrix}=\begin{pmatrix}
  \alpha_{C}& \alpha_{C}& \cdots &\alpha_{C}\\
  \alpha_{C}& \alpha_{C}& \cdots &\alpha_{C}\\
  \vdots  	& \vdots  & \ddots 	& \vdots\\
  \alpha_{C}& \alpha_{C}& \cdots &\alpha_{C}\\
 \end{pmatrix}
\end{align}

\begin{align}
\begin{pmatrix} \mathbf{A} & \mathbf{B} \\ \mathbf{C} & \mathbf{D} \end{pmatrix}^{-1}_{(n+m) \times (n+m)} = \begin{pmatrix}
  \alpha_{A}& \beta_{A}& \cdots &\beta_{A}		&\alpha_{B}& \alpha_{B}& \cdots &\alpha_{B}\\
  \beta_{A}& \alpha_{A}& \cdots &\beta_{A}		&  \alpha_{B}& \alpha_{B}& \cdots &\alpha_{B}\\
  \vdots  	& \vdots  & \ddots 	& \vdots			&  \vdots  	& \vdots  & \ddots 	& \vdots\\
  \beta_{A}& \beta_{A}& \cdots &\alpha_{A}		&  \alpha_{B}& \alpha_{B}& \cdots &\alpha_{B}\\
  \alpha_{C}& \alpha_{C}& \cdots &\alpha_{C}	&e& f& \cdots &f\\
  \alpha_{C}& \alpha_{C}& \cdots &\alpha_{C} &  f& e& \cdots &f\\
  \vdots  	& \vdots  & \ddots 	& \vdots			&  \vdots  	& \vdots  & \ddots 	& \vdots\\
  \alpha_{C}& \alpha_{C}& \cdots &\alpha_{C} &    f& f& \cdots &e
 \end{pmatrix}
\end{align}

\section{Finding Schur complements}\label{ap:schur}

\begin{align}
\mathbf{Q}_{2S\cdot 1S^c2S^c} &= \mathbf{Q}_{2S,2S} - \mathbf{Q}_{2S,1S^c2S^c} \mathbf{Q}^{-}_{1S^c2S^c,1S^c2S^c} \mathbf{Q}_{1S^c2S^c,2S}\\
&= \mathbf{Q}_{2S,2S} - \mathbf{Q}_{2S,1S^c2S^c}\begin{pmatrix} \mathbf{A} & \mathbf{B} \\ \mathbf{C} & \mathbf{D} \end{pmatrix}^{-1} \mathbf{Q}_{1S^c2S^c,2S}
\end{align}

where $\mathbf{Q}_{2S,2S}$ is $(N-m)\times(N-m)$ and  $\mathbf{Q}_{2S,1S^c2S^c}$ is $(N-m)\times(n+m)$. For $\mathbf{Q}_{2S,1S^c2S^c}$, the entries in the first $n$ columns are all $\rho_{12}$ and following $m$ columns are all $\rho_2$. We do matrix multiplication by partitioning.

\begin{align}
\mathbf{AB}=\begin{pmatrix} \mathbf{A}_{11} & \mathbf{A}_{12}\end{pmatrix}
\begin{pmatrix} \mathbf{B}_{11} & \mathbf{B}_{12} \\ \mathbf{B}_{21} & \mathbf{B}_{22} \end{pmatrix}
 = \begin{pmatrix} \mathbf{A}_{11}\mathbf{B}_{11} + \mathbf{A}_{12}\mathbf{B}_{21} & 
 \mathbf{A}_{11}\mathbf{B}_{21} + \mathbf{A}_{12}\mathbf{B}_{22} \end{pmatrix}
\end{align}

\begin{align}
\left(\mathbf{Q}_{2S,1S^c2S^c}\begin{pmatrix} \mathbf{A} & \mathbf{B} \\ \mathbf{C} & \mathbf{D} \end{pmatrix}^{-1}\right)_{(N-m) \times (n+m)}&=
\begin{pmatrix} {\boldsymbol\rho}_{12} & {\boldsymbol\rho}_{2}\end{pmatrix}
\begin{pmatrix} \boldsymbol{\alpha}_{A}\boldsymbol{\beta}_{A} & \boldsymbol{\alpha}_{B} \\ \boldsymbol{\alpha}_{C} & \mathbf{ef} \end{pmatrix}\\
 &=\begin{pmatrix} {\boldsymbol\rho}_{12}\boldsymbol{\alpha}_{A}\boldsymbol{\beta}_{A} + {\boldsymbol\rho}_{2}\boldsymbol{\alpha}_{C} & 
 {\boldsymbol\rho}_{12}\boldsymbol{\alpha}_{B} +  {\boldsymbol\rho}_{2}\mathbf{ef} \end{pmatrix}
 \end{align}
where ${\boldsymbol\rho}_{12}$ is $(N-m)\times n$ and ${\boldsymbol\rho}_{2}$ is $(N-m)\times m$.
 
\begin{align}
({\boldsymbol\rho}_{12}\boldsymbol{\alpha}_{A}\boldsymbol{\beta}_{A} )_{(N-m) \times n}
&=\begin{pmatrix}
  \rho_{12}& \rho_{12}& \cdots &\rho_{12}\\
  \rho_{12}& \rho_{12}& \cdots &\rho_{12}\\
  \vdots  	& \vdots  & \ddots 	& \vdots\\
    \rho_{12}& \rho_{12}& \cdots &\rho_{12}\\
 \end{pmatrix}_{(N-m) \times n} \times
 \begin{pmatrix}
  \alpha_{A}& \beta_{A}& \cdots &\beta_{A}		\\
  \beta_{A}& \alpha_{A}& \cdots &\beta_{A}		\\
  \vdots  	& \vdots  & \ddots 	& \vdots			\\
  \beta_{A}& \beta_{A}& \cdots &\alpha_{A}		
 \end{pmatrix}_{n \times n}\\
&= \begin{pmatrix}
  \rho_{12}(\alpha_{A}+ (n-1)\beta_{A})& \rho_{12}(\alpha_{A}+ (n-1)\beta_{A})& \cdots &\rho_{12}(\alpha_{A}+ (n-1)\beta_{A})\\
  \rho_{12}(\alpha_{A}+ (n-1)\beta_{A})& \rho_{12}(\alpha_{A}+ (n-1)\beta_{A})& \cdots &\rho_{12}(\alpha_{A}+ (n-1)\beta_{A})	\\
  \vdots  	& \vdots  & \ddots 	& \vdots			\\
  \rho_{12}(\alpha_{A}+ (n-1)\beta_{A})& \rho_{12}(\alpha_{A}+ (n-1)\beta_{A})& \cdots &\rho_{12}(\alpha_{A}+ (n-1)\beta_{A})
 \end{pmatrix}
\end{align}

\begin{align}
({\boldsymbol\rho}_{2}\boldsymbol{\alpha}_{C})_{(N-m) \times n}
&=\begin{pmatrix}
  m\rho_{2}\alpha_{C}& m\rho_{2}\alpha_{C}& \cdots &m\rho_{2}\alpha_{C}\\
  m\rho_{2}\alpha_{C}& m\rho_{2}\alpha_{C}& \cdots &m\rho_{2}\alpha_{C}\\
  \vdots  	& \vdots  & \ddots 	& \vdots\\
    m\rho_{2}\alpha_{C}& m\rho_{2}\alpha_{C}& \cdots &m\rho_{2}\alpha_{C}\\
 \end{pmatrix}
\end{align}

\begin{align}
({\boldsymbol\rho}_{12}\boldsymbol{\alpha}_{A}\boldsymbol{\beta}_{A} + {\boldsymbol\rho}_{2}\boldsymbol{\alpha}_{C})_{(N-m) \times n}
&=\begin{pmatrix}
   \rho_{12}(\alpha_{A}+ (n-1)\beta_{A})+m\rho_{2}\alpha_{C}& \cdots \\
  \vdots  	& \ddots 
 \end{pmatrix}
\end{align}

\begin{align}
({\boldsymbol\rho}_{12}\boldsymbol{\alpha}_{B})_{(N-m) \times m}
&=\begin{pmatrix}
  n\rho_{12}\alpha_{B}& n\rho_{12}\alpha_{B}& \cdots &n\rho_{12}\alpha_{B}\\
  n\rho_{12}\alpha_{B}&n\rho_{12}\alpha_{B}& \cdots &n\rho_{12}\alpha_{B}\\
  \vdots  	& \vdots  & \ddots 	& \vdots\\
   n\rho_{12}\alpha_{B}& n\rho_{12}\alpha_{B}& \cdots &n\rho_{12}\alpha_{B}\\
 \end{pmatrix}
\end{align}

\begin{align}
{\boldsymbol\rho}_{2}\mathbf{ef}_{(N-m) \times m}
&=\begin{pmatrix}
  \rho_{2}& \rho_{2}& \cdots &\rho_{2}\\
  \rho_{2}& \rho_{2}& \cdots &\rho_{2}\\
  \vdots  	& \vdots  & \ddots 	& \vdots\\
    \rho_{2}& \rho_{2}& \cdots &\rho_{2}\\
 \end{pmatrix}_{(N-m) \times m} \times
\begin{pmatrix}
  e& f& \cdots &f\\
  f& e& \cdots &f\\
  \vdots  	& \vdots  & \ddots 	& \vdots\\
    f& f& \cdots &e\\
 \end{pmatrix}_{m \times m}\\
&= \begin{pmatrix}
  \rho_{2}(e+ (m-1)f)& \rho_{2}(e+ (m-1)f)& \cdots &\rho_{2}(e+ (m-1)f)\\
  \rho_{2}(e+ (m-1)f)& \rho_{2}(e+ (m-1)f)& \cdots &\rho_{2}(e+ (m-1)f)	\\
  \vdots  	& \vdots  & \ddots 	& \vdots			\\
  \rho_{2}(e+ (m-1)f)& \rho_{2}(e+ (m-1)f)& \cdots &\rho_{2}(e+ (m-1)f)
 \end{pmatrix}
\end{align}

\begin{align}
 ({\boldsymbol\rho}_{12}\boldsymbol{\alpha}_{B} +  {\boldsymbol\rho}_{2}\mathbf{ef})_{(N-m) \times m}
 &=\begin{pmatrix}
   n\rho_{12}\alpha_{B}+\rho_{2}(e+ (m-1)f) & \cdots\\
  \vdots  	& \ddots 
 \end{pmatrix}
\end{align}

\begin{align}
\left(\mathbf{Q}_{2S,1S^c2S^c}\begin{pmatrix} \mathbf{A} & \mathbf{B} \\ \mathbf{C} & \mathbf{D} \end{pmatrix}^{-1}\right)_{(N-m) \times (n+m)} &=\begin{pmatrix} {\boldsymbol\rho}_{12}\boldsymbol{\alpha}_{A}\boldsymbol{\beta}_{A} + {\boldsymbol\rho}_{2}\boldsymbol{\alpha}_{C} & 
 {\boldsymbol\rho}_{12}\boldsymbol{\alpha}_{B} +  {\boldsymbol\rho}_{2}\mathbf{ef} \end{pmatrix}\\
 &=\begin{pmatrix}
   \rho_{12}(\alpha_{A}+ (n-1)\beta_{A})+m\rho_{2}\alpha_{C}& \cdots &n\rho_{12}\alpha_{B}+\rho_{2}(e+ (m-1)f) & \cdots\\
  \vdots  	& \ddots   &\vdots  	& \ddots 
 \end{pmatrix}
 \end{align}

\begin{align}
&\left(\mathbf{Q}_{2S,1S^c2S^c}\begin{pmatrix} \mathbf{A} & \mathbf{B} \\ \mathbf{C} & \mathbf{D} \end{pmatrix}^{-1} \mathbf{Q}_{1S^c2S^c,2S}\right)_{(N-m) \times (N-m)}\\
 &=\begin{pmatrix}
   \rho_{12}(\alpha_{A}+ (n-1)\beta_{A})+m\rho_{2}\alpha_{C}& \cdots &n\rho_{12}\alpha_{B}+\rho_{2}(e+ (m-1)f) & \cdots\\
  \vdots  	& \ddots   &\vdots  	& \ddots 
 \end{pmatrix}\begin{pmatrix}
  \rho_{12}& \rho_{12}& \cdots &\rho_{12}\\
  \rho_{12}& \rho_{12}& \cdots &\rho_{12}\\
  \vdots  	& \vdots  & \ddots 	& \vdots\\
  \rho_{12}& \rho_{12}& \cdots &\rho_{12}\\
  \rho_{2}& \rho_{2}& \cdots &\rho_{2}\\
  \rho_{2}& \rho_{2}& \cdots &\rho_{2}\\
  \vdots  	& \vdots  & \ddots 	& \vdots\\
  \rho_{2}& \rho_{2}& \cdots &\rho_{2}
 \end{pmatrix}_{(n+m) \times (N-m)}\\
  &=\begin{pmatrix}
   n\rho_{12}[\rho_{12}(\alpha_{A}+ (n-1)\beta_{A})+m\rho_{2}\alpha_{C}]+m\rho_{2}[n\rho_{12}\alpha_{B}+\rho_{2}(e+ (m-1)f)] & \cdots\\
  \vdots  	& \ddots  
 \end{pmatrix}\\
 &=\begin{pmatrix}
  \gamma& \gamma& \cdots &\gamma\\
  \gamma& \gamma& \cdots &\gamma\\
  \vdots  	& \vdots  & \ddots 	& \vdots\\
  \gamma& \gamma& \cdots &\gamma
 \end{pmatrix}
 \end{align}
 
 \begin{align}
(\mathbf{Q}_{2S\cdot 1S^c2S^c})_{(N-m) \times (N-m)}&= \mathbf{Q}_{2S,2S} - \mathbf{Q}_{2S,1S^c2S^c} \mathbf{Q}^{-}_{1S^c2S^c,1S^c2S^c} \mathbf{Q}_{1S^c2S^c,2S}\\
&=\begin{pmatrix}
  1& \rho_{2}& \cdots &\rho_{2}\\
  \rho_{2}& 1& \cdots &\rho_{2}\\
  \vdots  	& \vdots  & \ddots 	& \vdots\\
  \rho_{2}& \rho_{2}& \cdots &1
 \end{pmatrix}
- \begin{pmatrix}
  \gamma& \gamma& \cdots &\gamma\\
  \gamma& \gamma& \cdots &\gamma\\
  \vdots  	& \vdots  & \ddots 	& \vdots\\
  \gamma& \gamma& \cdots &\gamma
 \end{pmatrix}\\
&=\begin{pmatrix}
  1-\gamma& \rho_{2}-\gamma& \cdots &\rho_{2}-\gamma\\
  \rho_{2}-\gamma& 1-\gamma& \cdots &\rho_{2}-\gamma\\
  \vdots  	& \vdots  & \ddots 	& \vdots\\
  \rho_{2}-\gamma& \rho_{2}-\gamma& \cdots &1-\gamma
 \end{pmatrix}
\end{align}
\section{Finding expression for $\mathbf{1}^T \mathbf{Q}_{2S\cdot 1S^c2S^c}\mathbf{1}$}\label{ap:schurMutiplication}
\begin{align*}
\mathbf{1}^T \mathbf{Q}_{2S\cdot 1S^c2S^c}\mathbf{1}&=
(N-m)[(1-\gamma)+(N-m-1)(\rho_{2}-\gamma)]\\
&=(N-m)[1+(N-m-1)\rho_{2}-(N-m)\gamma]\\
&\triangleq \phi(N-n,M-m,\rho_{1,2,12})
 \end{align*}
\begin{remark}
$\mathbf{1}^T \mathbf{Q}_{2S\cdot 1S^c2S^c}\mathbf{1}$ is by definition either $\phi(N-n,M-m,\rho_{1,2,12})$ or $\phi(n,m,\rho_{1,2,12})$ depending on either $|1S|=N-n,|2S|=N-m$ or $|1S|=n,|2S|=m$.
\end{remark}
Plugging in $\gamma$,
\begin{align*}
&\phi(N-n,M-m,\rho_{1,2,12})\\
&=(N-m)[1+(N-m-1)\rho_{2}-(N-m)n\rho_{12}[\rho_{12}(\alpha_{A}+ (n-1)\beta_{A})+m\rho_{2}\alpha_{C}]\\
&{ \ \  \ \ }-(N-m)m\rho_{2}[n\rho_{12}\alpha_{B}+\rho_{2}(e+ (m-1)f)] ]
 \end{align*} 
Letting $N-m=l$
\begin{align*}
&\phi(N-n,M-m,\rho_{1,2,12})\\
&=l[1+(l-1)\rho_{2}-ln\rho_{12}[\rho_{12}(\alpha_{A}+ (n-1)\beta_{A})+m\rho_{2}\alpha_{C}]-lm\rho_{2}[n\rho_{12}\alpha_{B}+\rho_{2}(e+ (m-1)f)] ]\\
&=l[1+(l-1)\rho_2 - ln\rho_{12}^2 \alpha_{A}\\
&\hspace{2.65cm}-ln(n-1) \rho_{12}^2 \beta_{A}\\
&\hspace{2.65cm}- lmn\rho_2 \rho_{12}\alpha_{C} \\
&\hspace{2.65cm}-lmn\rho_2 \rho_{12} \alpha_{B}\\
&\hspace{2.65cm}-lm\rho_{2}^2(e + (m-1)f) ]\\
&=l[1+(l-1)\rho_2 - ln\rho_{12}^2 [x+w(x+(n-1)y)^2]\\
&\hspace{2.65cm}-ln(n-1) \rho_{12}^2 [y+w(x+(n-1)y)^2]\\
&\hspace{2.65cm}+lmn\rho_2 \rho_{12}((m-1)fi+ei) \\
&\hspace{2.65cm}+lmn\rho_2 \rho_{12} ((n-1)yh+xh)	\\
&\hspace{2.65cm}-lm\rho_{2}^2(e + (m-1)f) ]\\
&=l[1+(l-1)\rho_2 - ln\rho_{12}^2 [x+w(x+(n-1)y)^2]\\
&\hspace{2.65cm}-ln(n-1) \rho_{12}^2 [y+w(x+(n-1)y)^2]\\
&\hspace{2.65cm}+ lmn\rho_2 \rho_{12}((m-1)f+e)\rho_{12}((n-1)y+x) \\
&\hspace{2.65cm}+lmn\rho_2 \rho_{12} ((n-1)y+x)\rho_{12}((m-1)f+e)	\\
&\hspace{2.65cm}-lm\rho_{2}^2(e + (m-1)f) ]\\
&=l[1+(l-1)\rho_2 - ln\rho_{12}^2 [x+w(x+(n-1)y)^2]\\
&\hspace{2.65cm}-ln(n-1) \rho_{12}^2 [y+w(x+(n-1)y)^2]\\
&\hspace{2.65cm}+ lmn\rho_2 \rho_{12}^2(e+(m-1)f)(x+(n-1)y) \\
&\hspace{2.65cm}+lmn\rho_2 \rho_{12}^2(x+(n-1)y)(e+(m-1)f)	\\
&\hspace{2.65cm}-lm\rho_{2}^2(e + (m-1)f) ]\\
&=l[1+(l-1)\rho_2 - ln\rho_{12}^2 [x+m \rho_{12}^2 (e+ (m-1)f)(x+(n-1)y)^2]\\
&\hspace{2.65cm}-ln(n-1) \rho_{12}^2 [y+m\rho_{12}^2 (e+ (m-1)f)(x+(n-1)y)^2]\\
&\hspace{2.65cm}+lmn\rho_2 \rho_{12}^2(e+(m-1)f)(x+(n-1)y) \\
&\hspace{2.65cm}+lmn\rho_2 \rho_{12}^2(x+(n-1)y)(e+(m-1)f)	\\
&\hspace{2.65cm}-lm\rho_{2}^2(e + (m-1)f) ]\\
&=l[1+(l-1)\rho_2 - ln\rho_{12}^2 [x+m \rho_{12}^2 (e+ (m-1)f)(x+(n-1)y)^2]\\
&\hspace{2.65cm}-ln(n-1) \rho_{12}^2 [y+m\rho_{12}^2 (e+ (m-1)f)(x+(n-1)y)^2]\\
&\hspace{2.65cm}+2lmn\rho_2 \rho_{12}^2(e+(m-1)f)(x+(n-1)y) \\
&\hspace{2.65cm}-lm\rho_{2}^2(e + (m-1)f) ]\\
&=l[1+(l-1)\rho_2 - ln\rho_{12}^2 x\\
&\hspace{2.65cm} - lnm \rho_{12}^4 (e+ (m-1)f)(x+(n-1)y)^2\\
&\hspace{2.65cm}-ln(n-1) \rho_{12}^2 y\\
&\hspace{2.65cm}-lnm(n-1)\rho_{12}^4 (e+ (m-1)f)(x+(n-1)y)^2\\
&\hspace{2.65cm}+2lmn\rho_2 \rho_{12}^2(e+(m-1)f)(x+(n-1)y) \\
&\hspace{2.65cm}-lm\rho_{2}^2(e + (m-1)f) ]\\
&=l[1+(l-1)\rho_2 - ln\rho_{12}^2 (x+(n-1)y)\\
&\hspace{2.65cm} - ln^2m \rho_{12}^4 (e+ (m-1)f)(x+(n-1)y)^2\\
&\hspace{2.65cm}+2lmn\rho_2 \rho_{12}^2(e+(m-1)f)(x+(n-1)y) \\
&\hspace{2.65cm}-lm\rho_{2}^2(e + (m-1)f) ]\\
&=l[1+(l-1)\rho_2 - ln\rho_{12}^2 (x+(n-1)y)\\
&\hspace{2.65cm} - ln^2m \rho_{12}^4 (e+ (m-1)f)(x+(n-1)y)^2\\
&\hspace{2.65cm}+2lmn\rho_2 \rho_{12}^2(e+(m-1)f)(x+(n-1)y) \\
&\hspace{2.65cm}-lm\rho_{2}^2(e + (m-1)f) ]\\
%
%
\end{align*}
where
\begin{align*}
x&=\frac{(n-2)\rho_1+1}{-(n-1)\rho_1^2+(n-2)\rho_1+1}\\
y&= \frac{-\rho_1}{-(n-1)\rho_1^2+(n-2)\rho_1+1}\\
x+(n-1)y&= \frac{1}{1+(n-1)\rho_1}
\end{align*}

 \begin{align*}
&e+(m-1)f\\
&=\frac{(m-2)\rho_2 +1-(m-1)n\rho_{12}^2 (x+(n-1)y)-(m-1)[\rho_2-n\rho_{12}^{2}(x+(n-1)y)]}{1+(m-2)\rho_2-(m-1)\rho_2^2+mn(\rho_2-1)\rho_{12}^2(x+(n-1)y)}\\
&=\frac{(m-2)\rho_2 +1-(m-1)n\rho_{12}^2 (x+(n-1)y)-(m-1)\rho_2+(m-1)n\rho_{12}^{2}(x+(n-1)y)}{1+(m-2)\rho_2-(m-1)\rho_2^2+mn(\rho_2-1)\rho_{12}^2(x+(n-1)y)}\\
&=\frac{1-\rho_2}{1+(m-2)\rho_2-(m-1)\rho_2^2+mn(\rho_2-1)\rho_{12}^2(x+(n-1)y)}\\
&=\frac{1-\rho_2}{(1-\rho_2)[1+(m-1)\rho_2]-mn(1-\rho_2)\rho_{12}^2(x+(n-1)y)}\\
&=\frac{1}{1+(m-1)\rho_2-mn\rho_{12}^2(x+(n-1)y)}\\
&=\frac{1}{1+(m-1)\rho_2-mn\rho_{12}^2\frac{1}{1+(n-1)\rho_1}}\\
&=\frac{[1+(n-1)\rho_1]}{[1+(n-1)\rho_1][1+(m-1)\rho_2]-mn\rho_{12}^2}\\
\end{align*}

\begin{align*}
&e\\
&=\frac{(m-2)v+u}{-(m-1)v^2+(m-2)uv+u^2} 	\\
&=\frac{(m-2)(\rho_2 -z)+(1-z)}{-(m-1)(\rho_2-z)^2+(m-2)(\rho_2-z)(1-z)+(1-z)^2}\\
&=\frac{1+(m-2)\rho_2 -(m-1)z}{1+(m-2)\rho_2-(m-1)\rho_2^2+(m\rho_2-m)z}\\
&=\frac{(m-2)\rho_2 +1 -(m-1)[n\rho_{12} ((n-1)\rho_{12}y+\rho_{12}x)]}{{1+(m-2)\rho_2-(m-1)\rho_2^2+(m\rho_2-m)[n\rho_{12} ((n-1)\rho_{12}y+\rho_{12}x)]}}\\
&=\frac{(m-2)\rho_2 +1 -(m-1)n\rho_{12}^2 (x+(n-1)y)}{1+(m-2)\rho_2-(m-1)\rho_2^2+mn(\rho_2-1)\rho_{12}^2(x+(n-1)y)}\\
\end{align*}
Where the denominator term for $e$ and $f$ is derived as follows.
\begin{align*}
   &-(m-1)(\rho_2-z)^2+(m-2)(\rho_2-z)(1-z)+(1-z)^2\\
=&-(m-1)\rho_2^2+ (m-1)2\rho_2z-(m-1)z^2+(m-2)(\rho_2-\rho_2z-z+z^2)+1-2z+z^2\\
=&-(m-1)\rho_2^2+ (m-1)2\rho_2z-(m-1)z^2+(m-2)\rho_2-(m-2)\rho_2z-(m-2)z+(m-2)z^2+1-2z+z^2\\
=&-(m-1)\rho_2^2+ (m-1)2\rho_2z+(m-2)\rho_2-(m-2)\rho_2z-(m-2)z+1-2z\\
=&1+(m-2)\rho_2-(m-1)\rho_2^2+ (m-1)2\rho_2z-(m-2)\rho_2z-mz\\
=&1+(m-2)\rho_2-(m-1)\rho_2^2+ (m-1)2\rho_2z-(m-1)\rho_2z+\rho_2z-mz\\
=&1+(m-2)\rho_2-(m-1)\rho_2^2+ (m-1)\rho_2z+\rho_2z-mz\\
=&1+(m-2)\rho_2-(m-1)\rho_2^2+ m\rho_2z-mz\\
=&1+(m-2)\rho_2-(m-1)\rho_2^2+ (m\rho_2-m)z\\
\end{align*}

\begin{align*}
f&= -\frac{v}{-(m-1)v^2+(m-2)uv+u^2} \\
&=-\frac{\rho_2-z}{-(m-1)(\rho_2-z)^2+(m-2)(1-z)(\rho_2-z)+(1-z)^2} \\
&=-\frac{\rho_2-z}{1+(m-2)\rho_2-(m-1)\rho_2^2+(m\rho_2-m)z} \\
&=-\frac{\rho_2-[n\rho_{12} ((n-1)\rho_{12}y+\rho_{12}x)]}{1+(m-2)\rho_2-(m-1)\rho_2^2+(m\rho_2-m)[n\rho_{12} ((n-1)\rho_{12}y+\rho_{12}x)]} \\
&=-\frac{\rho_2-n\rho_{12}^2 (x+(n-1)y)}{1+(m-2)\rho_2-(m-1)\rho_2^2+mn(\rho_2-1)\rho_{12}^2(x+(n-1)y)} \\
 \end{align*}
 
 \begin{align*}
&\phi(N-n,N-m,\rho_{1,2,12})\\
=&l[1+(l-1)\rho_2 \\
&- ln\rho_{12}^2 (x+(n-1)y)\\
&- ln^2m \rho_{12}^4 (e+ (m-1)f)(x+(n-1)y)^2\\
&+2lmn\rho_2 \rho_{12}^2(e+(m-1)f)(x+(n-1)y) \\
&-lm\rho_{2}^2(e + (m-1)f) ]\\
=&(N-m)[1+(N-m-1)\rho_2 \\
&- (N-m)n\rho_{12}^2 \frac{1}{1+(n-1)\rho_1}\\
&- ln^2m \rho_{12}^4 \left(\frac{[1+(n-1)\rho_1]}{[1+(n-1)\rho_1][1+(m-1)\rho_2]-mn\rho_{12}^2}\right)\left(\frac{1}{1+(n-1)\rho_1}\right)^2\\
&+2lmn\rho_2 \rho_{12}^2\left(\frac{1}{[1+(n-1)\rho_1][1+(m-1)\rho_2]-mn\rho_{12}^2}\right) \\
&-lm\rho_{2}^2\left(\frac{[1+(n-1)\rho_1]}{[1+(n-1)\rho_1][1+(m-1)\rho_2]-mn\rho_{12}^2}\right) ]\\
=&(N-m)[1+(N-m-1)\rho_2 \\
&- (N-m)n\rho_{12}^2 \frac{1}{1+(n-1)\rho_1}\\
&- (N-m)n^2m \rho_{12}^4 \left(\frac{1}{[1+(n-1)\rho_1]^2[1+(m-1)\rho_2]-mn\rho_{12}^2[1+(n-1)\rho_1]}\right)\\
&+2(N-m)mn\rho_2 \rho_{12}^2\left(\frac{1}{[1+(n-1)\rho_1][1+(m-1)\rho_2]-mn\rho_{12}^2}\right) \\
&-(N-m)m\rho_{2}^2\left(\frac{[1+(n-1)\rho_1]}{[1+(n-1)\rho_1][1+(m-1)\rho_2]-mn\rho_{12}^2}\right) ]\\
\end{align*}
Letting (or labeling) $m$ as $N-m$ and $n$ as $N-n$ (note that, this does not change the set of bounds for different cuts)
\begin{align*}
&\phi(n,m,\rho_{1,2,12})\\
=&m[1+(m-1)\rho_2 \\
&- \frac{m(N-n)\rho_{12}^2}{1+(N-n-1)\rho_1}\\
&- \frac{m(N-n)^2(N-m) \rho_{12}^4}{[1+(N-n-1)\rho_1]} \left(\frac{1}{[1+(N-n-1)\rho_1][1+(N-m-1)\rho_2]-(N-m)(N-n)\rho_{12}^2}\right)\\
&+2m(N-m)(N-n)\rho_2 \rho_{12}^2\left(\frac{1}{[1+(N-n-1)\rho_1][1+(N-m-1)\rho_2]-(N-m)(N-n)\rho_{12}^2}\right) \\
&-m(N-m)\rho_{2}^2\left(\frac{[1+(N-n-1)\rho_1]}{[1+(N-n-1)\rho_1][1+(N-m-1)\rho_2]-(N-m)(N-n)\rho_{12}^2}\right) ]\\
=&m[1+(m-1)\rho_2 \\
&- \frac{m(N-n)\rho_{12}^2}{1+(N-n-1)\rho_1}\\
&- \frac{m(N-n)^2(N-m) \rho_{12}^4}{[1+(N-n-1)\rho_1]\left([1+(N-n-1)\rho_1][1+(N-m-1)\rho_2]-(N-m)(N-n)\rho_{12}^2\right)}\\
&+\frac{2m(N-m)(N-n)\rho_2 \rho_{12}^2}{[1+(N-n-1)\rho_1][1+(N-m-1)\rho_2]-(N-m)(N-n)\rho_{12}^2} \\
&-\frac{m(N-m)\rho_{2}^2[1+(N-n-1)\rho_1]}{[1+(N-n-1)\rho_1][1+(N-m-1)\rho_2]-(N-m)(N-n)\rho_{12}^2} ]\\
\end{align*}
Alternatively,
\begin{align*}
\phi(n,m,\rho_{1,2,{12}})&=m\big(1+(m-1)\rho_2 - m(N-n)\rho_{12}^2 (x+(N-n-1)y)\\
&\text{ \ \ \ \  \ \ }- m(N-n)^2(N-m) \rho_{12}^4 (e+ (N-m-1)f)(x+(N-n-1)y)^2\\
&\text{ \ \ \ \  \ \ }+2m(N-m)(N-n)\rho_2 \rho_{12}^2(e+(N-m-1)f)(x+(N-n-1)y) \\
&\text{ \ \ \ \  \ \ }-m(N-m)\rho_{2}^2(e + (N-m-1)f) \big)\\
\text{where }x+(N-n-1)y&= \frac{1}{1+(N-n-1)\rho_1}\\
\text{and } e+(N-m-1)f
&=\frac{(1+(N-n-1)\rho_1)}{(1+(N-n-1)\rho_1)(1+(N-m-1)\rho_2)-(N-m)(N-n)\rho_{12}^2}.
\end{align*}
\section{$\rho_{12}$ is a critical point and a maxima}\label{ap:criticalNmaxima}
 \begin{align*}
\phi(\rho_1,\rho_2,\rho_{12})
=&l[1+(l-1)\rho_2 \\
&- ln\rho_{12}^2 (x+(n-1)y)\\
&- ln^2m \rho_{12}^4 (e+ (m-1)f)(x+(n-1)y)^2\\
&+2lmn\rho_2 \rho_{12}^2(e+(m-1)f)(x+(n-1)y) \\
&-lm\rho_{2}^2(e + (m-1)f) ]\\
\phi'(\rho_{12})=
&l[-2ln\rho_{12} (x+(n-1)y)\\
&- 4ln^2m \rho_{12}^3 (e+ (m-1)f)(x+(n-1)y)^2\\
&- ln^2m \rho_{12}^4 \frac{d}{d \rho_{12}}((e+ (m-1)f)(x+(n-1)y)^2)\\
&+4lmn\rho_2 \rho_{12}(e+(m-1)f)(x+(n-1)y)	\\
&+2lmn\rho_2 \rho_{12}^2 \frac{d}{d \rho_{12}}((e+(m-1)f)(x+(n-1)y))\\
&-0 - lm\rho_{2}^2\frac{d}{d \rho_{12}}((e + (m-1)f)) ]
 \end{align*}

\begin{align*}
\frac{d}{d \rho_{12}}(e+(m-1)f)&=\frac{+[1+(n-1)\rho_1] 2mn\rho_{12}}{[[1+(n-1)\rho_1][1+(m-1)\rho_2]-mn\rho_{12}^2]^2}=\frac{N}{D}\\
&= 2mn \rho_{12} (e+(m-1)f)^2 (x+(n-1)y)\\
{\frac{d}{d \rho_{12}}(e+(m-1)f)}_{|\rho_{12}=0}&=0\\
\end{align*}

Note that $\phi'(\rho_{12}=0)=0$ and hence $\rho_{12}=0$ is a critical point. Now,
 \begin{align*}
\phi''(\rho_{12}=0)=&l[-2ln(x+(n-1)y)\\
&+4lmn\rho_2 (x+(n-1)y) \frac{d}{d \rho_{12}}(e+(m-1)f)	\\
&- lm\rho_{2}^2\frac{d^2}{d \rho_{12}^2}(e + (m-1)f)\\
=&l\left[-2ln(x+(n-1)y)- lm\rho_{2}^2\frac{2nm}{[1+(n-1)\rho_1][1+(m-1)\rho_2]^2}\right]\\
=&2l^2n (x+(n-1)y)\left[-1- \frac{m^2\rho_{2}^2}{[1+(m-1)\rho_2]^2}\right]
 \end{align*}
Where
\begin{align*}
{\frac{d^2}{d \rho_{12}^2}(e+(m-1)f)}_{|\rho_{12}=0}
=\frac{2mn}{[1+(n-1)\rho_1][1+(m-1)\rho_2]^2}
\end{align*}

Note that the term multiplying $x +(n-1)y$ is negative. Also, for $\rho_{12}=0$, we have $\rho_{1}\geq \frac{-1}{N-1}$ and hence $x +(n-1)y$ is non-negative for $\rho_1 \in [\frac{-1}{N-1},1]$. Thus, $\phi''(\rho_{12}=0)$ is negative and hence $\rho_{12}=0$ is a 
maxima.



\end{appendices}

\end{document}